\documentclass{article}
\usepackage[margin=0.8in]{geometry}
\usepackage{amsmath,amsthm,amssymb,amsfonts,bm, xcolor}
\usepackage{setspace,hyperref,url,graphicx,multirow}

\usepackage{soul}
\usepackage{authblk}
\usepackage{enumitem}
\usepackage{algorithm,algpseudocode}
\usepackage[sorting = none, backend = biber, style=numeric-comp]{biblatex}

\providecommand{\keywords}[1]
{
  \small	
  \textbf{\textit{Keywords---}} #1
}

\title{A Robust Cybersecurity Topic Classification Tool}

\author[1]{Elijah Pelofske\thanks{Email: elijahpelofske@gmail.com}}
\author[1]{Lorie M. Liebrock}
\author[2]{Vincent Urias}
\affil[1]{New Mexico Cybersecurity Center of Excellence, New Mexico Institute of Mining and Technology, Socorro, New Mexico, USA}
\affil[2]{Sandia National Laboratories, Albuquerque, New Mexico, USA}

\begin{document}

\date{\vspace{-5ex}}

\maketitle

\begin{abstract}
\noindent In this research, we use user defined labels from three internet text sources (Reddit, StackExchange, arXiv) to train 21 different machine learning models for the topic classification task of detecting cybersecurity discussions in natural English text. We analyze the false positive and false negative rates of each of the 21 model's in cross validation experiments. Then we present a Cybersecurity Topic Classification (CTC) tool, which takes the majority vote of the 21 trained machine learning models as the decision mechanism for detecting cybersecurity related text. We also show that the majority vote mechanism of the CTC tool provides lower false negative and false positive rates on average than any of the 21 individual models. We show that the CTC tool is scalable to the hundreds of thousands of documents with a wall clock time on the order of hours. 
\end{abstract}


\section{Introduction}
\label{sec:introduction}

Identifying cybersecurity discussions in open forums at scale is a topic of great interest for the purpose of mitigating and understanding modern cyber threats \cite{hughes-etal-2020-detecting, DBLP:conf/aaai/LippmanWMCCS017, minaee2021deep}. The challenge is that these discussions are typically quite noisy (i.e., they contain community known synonyms or acronyms or slang) and it is difficult to get labelled data in order to train resilient NLP (natural language processing) topic classifiers. Additionally, it is important that a tool that detects cybersecurity discussions in internet text sources is \textit{scalable} and offers \textit{low errors rates} (in particular, both low false negative rates and low false positive rates). 

In order to address the challenges of finding relevant cybersecurity labelled data, we use a technique that gathers posts or articles from different internet sources that have user defined \emph{topic labels}. We then collect and label the training text as being cybersecurity related or not based on the subset of labels that the text source offers. Thus, the labelled training data we gather is not manually labelled by researchers; instead it is labelled inherently by the system we gather the text from. This provides an additional benefit  for cybersecurity related discussions in that it might be difficult for a manual labelling process to catch all of the language variation (e.g., unknown synonyms). This labelling method uses the user defined labels (which removes the need for the labelling process to identify all known cybersecurity terms used in online discussion forums) and provides a much larger and current labelled data set. Lastly, this method of gathering labelled data is \emph{highly} scalable. The reason is that the platforms we used have publicly available data and systems to retrieve that data in very large amounts. We used three specific sources of text: Reddit, StackExchange, and Arxiv. 

Using the topic classification labelled data, we train a total of 21 different machine learning models using several different algorithms and the three different text sources. We then show the validation accuracy of the models, as well as the cross validation (i.e., validating a model on a text source upon which it was not trained) accuracy of each of the models. After that we define a confidence measure for the continuous output values of the machine learning models and present those results for the models where applicable. We next present the Cybersecurity Topic Classification (CTC) tool, which uses the majority vote consensus of all 21 trained models in order to evaluate whether a novel document is cybersecurity related or not. We show that the CTC tool has both scalability to hundreds of thousands of documents per hour and low error rates. Lastly, we provide all of the labelled data we used to train and validate the models in a \href{https://github.com/epelofske-student/CTC}{Github repository} \cite{CTC-github} and a Zenodo dataset \cite{pelofske_2024_10655913}.

This article is structured as follows. After a brief literature review in Section \ref{sec:previous_work}, we define our methods of gathering labelled data and then training and validating the machine learning models in Section \ref{sec:methods}. Section \ref{sec:experiments} describes the experiments. The investigation of how the minimum token length of the training data changes false negative and false positive rates is presented in Section \ref{sec:experiments_token_length}. After training each of the models using the specified parameters, in Section \ref{sec:experiments_validation} we show the validation accuracy rates for each of the 21 trained models. In Section \ref{sec:experiments_confidence} we show that the continuous outputs of some of the machine learning models can yield a simple confidence measure, where higher confidence means higher solution quality. In Section \ref{sec:CTC} we present the CTC tool and show it's scalability (meaning real time text samples classified per hour) and high accuracy. Section \ref{sec:conclusion} discusses conclusions and future work. 

\subsection{Previous work}
\label{sec:previous_work}

There is significant interest surrounding the goal of being able to automate cybersecurity threat detection on social media \cite{DBLP:conf/aaai/LippmanWMCCS017, DBLP:journals/corr/abs-2008-04176, 10.1145/3132847.3132866, sentiment-twitter, hughes-etal-2020-detecting, 10.1007/978-3-319-93372-6_41, Ikwu_2019, DeepLearningDDosTwitter}. Twitter, Reddit, and Stackexchange are popular forums from which several previous studies have gathered cybersecurity related documents \cite{DBLP:conf/aaai/LippmanWMCCS017, sentiment-twitter, Ikwu_2019, 10.1145/3132847.3132866, DeepLearningDDosTwitter, oro59243} for the purpose of training machine learning detection systems and classifiers. In particular, \cite{DBLP:conf/aaai/LippmanWMCCS017} used tags (or other community defined mechanisms) as a document labelling method for cybersecurity topic classification related text. There is also some interest in investigating vulnerability discussions on developer sites such as Stackoverflow \cite{DBLP:journals/corr/abs-2008-04176, oro59243}. 

There are several different approaches taken with which topic modelling task to use as a signal to detect cybersecurity discussions. Typically the topic classification task is related to training directly on labelled text and then perhaps developing an idea of the more relevant keywords in these discussions \cite{DBLP:conf/aaai/LippmanWMCCS017, hughes-etal-2020-detecting}. Other researchers use sentiment analysis in conjunction with machine learning models \cite{10.1007/978-3-319-93372-6_41, sentiment-twitter}. 

There are also interesting approaches that use social media as a signal to detect specific cyber-attacks (e.g., DDoS attacks) or vulnerabilities \cite{10.1145/3132847.3132866, DBLP:journals/corr/abs-2008-04176, DBLP:conf/aaai/LippmanWMCCS017, DeepLearningDDosTwitter}. 

For a review of text classification with deep learning models, see \cite{minaee2021deep}, and for a survey of gathering social media data, see \cite{Batrinca2015}. 

\section{Methods}
\label{sec:methods}

This section describes the methods used to gather labelled text, preprocess and vectorize that text, train several machine learning models using the labelled data, and lastly evaluate the accuracy of these machine learning models. 

All figures in this article were created using Matplotlib \cite{Hunter:2007}. 

\subsection{Text sources}
\label{sec:methods_text_sources}
For this research, we focus on gathering large amounts of text from the three sources Reddit, Stackexchange, and Arxiv. For gathering Reddit text we used the python modules \textit{praw} and \textit{psaw} \cite{praw, psaw}. In order to query data from StackExchange, we used the python module \textit{StackAPI} (which is a python wrapper for the StackExchange API \cite{StackAPI}). For gathering arXiv documents, we used the python module \textit{arxiv} \cite{arxivAPI}. As a summary of the raw data collected, Table \ref{table:text_source_data} shows the number of cybersecurity and non cybersecurity documents gathered from each source, along with the document labelling method for each source. Next we define the precise methodology for gathering and labelling the documents from each source.

\begin{table}[t]
\caption{Number of documents from each text source and the labelling method used for each source}
\label{table:text_source_data}
\begin{center}
\begin{tabular}{ | l | l | l | l | }
\hline
Source & cybersecurity & non cybersecurity & labelling method \\
\hline
\hline
Reddit & 164750 & 4184184 & sub-reddits \\ 
\hline
StackExchange & 41162 & 4842461 & post topic labels \\ 
\hline
Arxiv & 12132 & 28996 & keyword search + topic restriction \\ 
\hline
\end{tabular}
\end{center}
\end{table}

\subsubsection{Reddit}
\label{sec:methods_text_sources_reddit}

Reddit \cite{reddit} is an internet discussion website that allows registered users to submit different types of content (e.g., text posts, images, links, videos) to the site. Each of these posts then get voted on (upvoted or downvoted) by other users. The entire site is logically organized into sub-reddits. Each sub-reddit has a specific scope and topic of discussion. The document labeling strategy we use is to label documents according to which sub-reddit they originate from. 

For gathering cybersecurity related text, we first defined 40 cybersecurity related sub-reddits. Next, we queried each of those sub-reddits with a maximum number of posts returned of $1,000,000$ (This does not mean we get $1,000,000$ posts for each sub-reddit). The default sorting method for the posts returned is a Reddit defined post metric called \textit{Hot}; for the purpose of querying with the API, this can not be changed. 

For gathering non cybersecurity related text, we search the top 100 most popular sub-reddits at the time of searching (this list can also be found at \cite{CTC-github}) using the same method described above. None of the 40 cybersecurity topic focused sub-reddits are in the top 100 most popular sub-reddits.  

For gathering posts in general, we perform some filtering of the data before actually labelling and storing each document. In particular, we remove all posts which are marked as \textbf{deleted} or \textbf{removed}, since those posts do not contain any post text anymore (these posts were either removed by the user who posted it or a Reddit administrator). 

For each post, the title and main post are treated separately in the API. In order to construct a document out of each post, we treated the title as the first sentence and the post content as the remainder of the document (i.e., we merged the two pieces of text with a period and space in between). 

\begin{figure}[h!]
    \centering
    \includegraphics[width=0.49\textwidth]{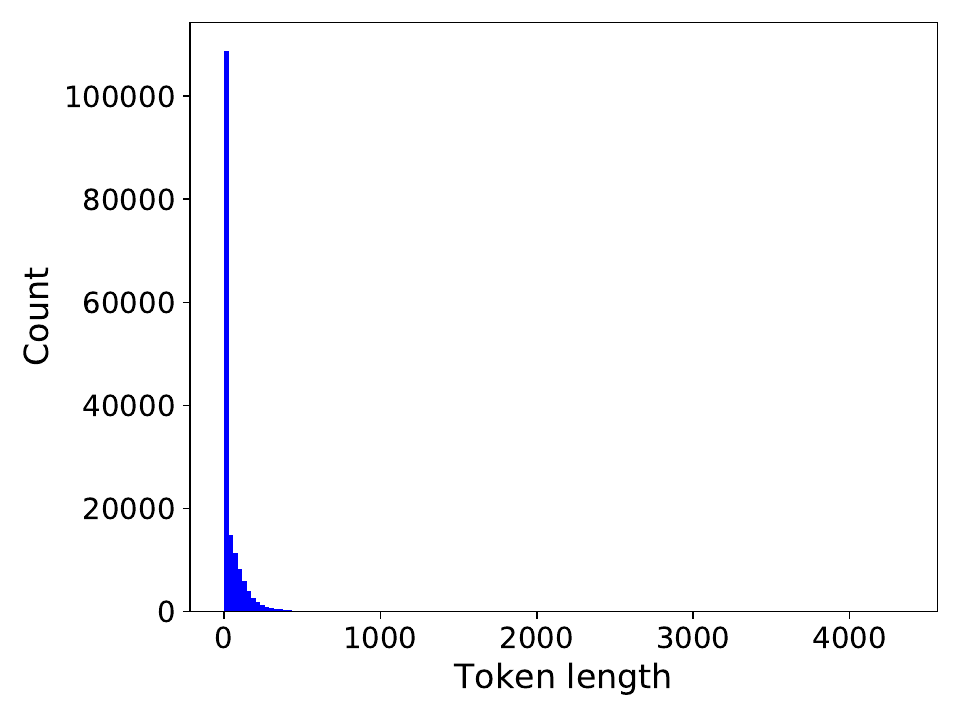}
    \includegraphics[width=0.49\textwidth]{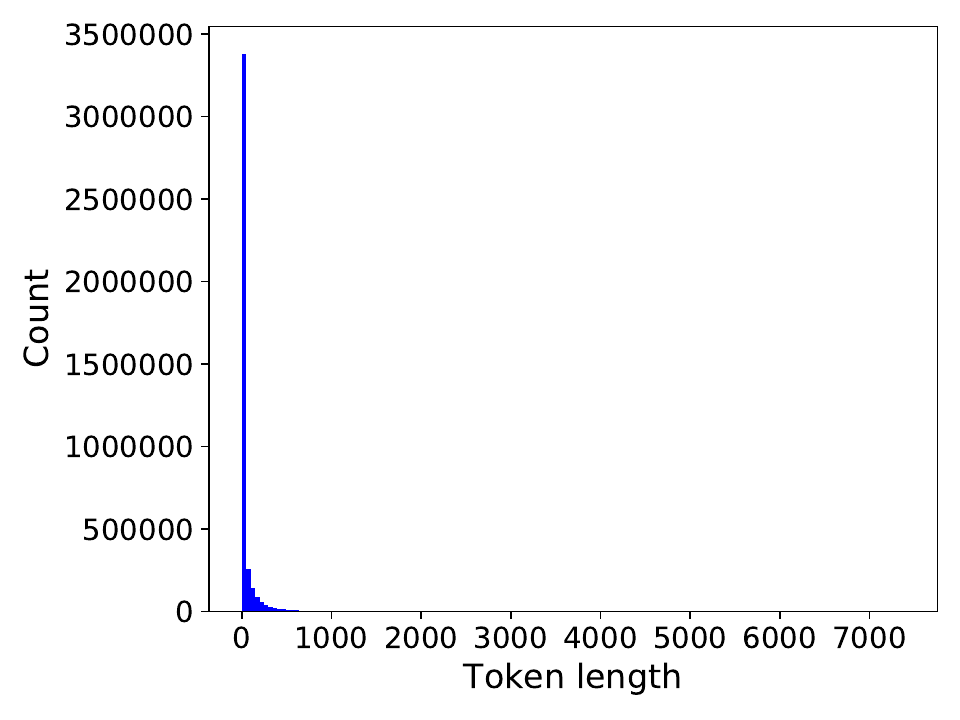}
    \caption{Reddit token data histogram; Cybersecurity documents (left) and non-cybersecurity documents (right). }
    \label{fig:Reddit_token_length_histogram}
\end{figure}

Figure \ref{fig:Reddit_token_length_histogram} shows the distribution of the collected tagged Reddit text in terms of token (i.e., usable words) length.

\subsubsection{StackExchange}
\label{sec:methods_text_sources_stackexchange}

StackExchange \cite{stackexchange} is a group of Q\&A websites whose topics of discussion are wide ranging; but the most popular sites are developer and programming sites such as Stackoverflow. The sites are self moderating in that registered user's can upvote and downvote posts. Each site also allows the users who post to identify posts using topic tags. 

We used multiple StackExchange sites as text sources (see \cite{CTC-github} for the full list). For each of these StackExchange sites, we gathered the top (defined by most upvoted) 10,000 posts for each month since the inception of the given StackExchange site. 

Next, we defined a list of cybersecurity related topic tags across different security and technology related StackExchange sites (this list can be found at our Github \cite{CTC-github}). We labelled each post as being cybersecurity related if it used any of the tags in our list and otherwise it was labelled as not cybersecurity related. 

As with Reddit, for each post we queried, we merged the title and main post text into a piece of text. 

\begin{figure}[h!]
    \centering
    \includegraphics[width=0.49\textwidth]{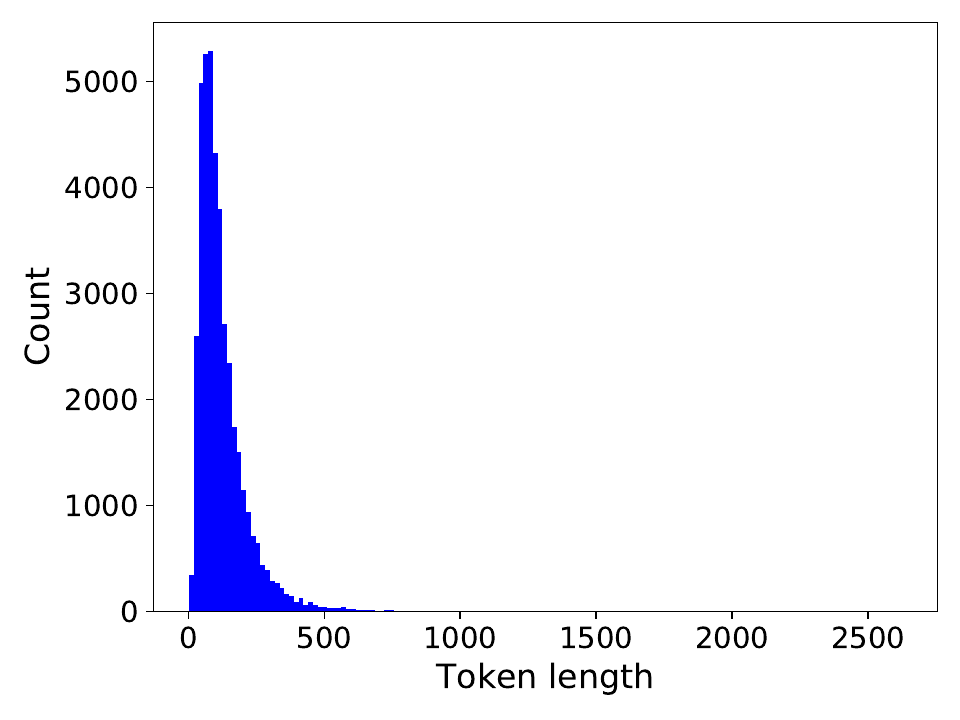}
    \includegraphics[width=0.49\textwidth]{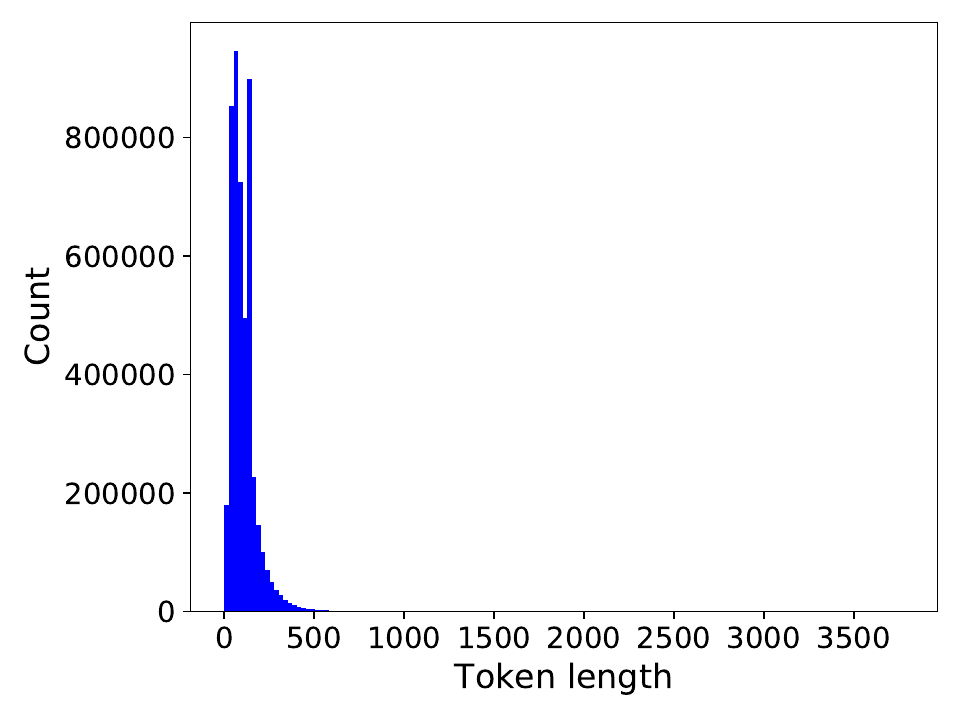}
    \caption{StackExchange token data histogram; Cybersecurity documents (left) and non-cybersecurity documents (right). }
    \label{fig:Stackexchange_token_length_histogram}
\end{figure}

Figure \ref{fig:Stackexchange_token_length_histogram} shows the distribution of the collected tagged StackExchange text in terms of token (i.e., usable parsed word) length.

\subsubsection{Arxiv}
\label{sec:methods_text_sources_arxiv}

Arxiv \cite{arxiv} is an open access repository of e-prints (including papers before peer-review and after peer-review). The repository includes scientific papers on wide ranging topics including computer science, mathematics, physics, statistics, and economics. Each paper comes with one more topic labels and can be downloaded in the form of a PDF. 

The methodology for gathering cybersecurity labelled text is as follows. We used a seed list of cybersecurity terms and topics in order to search arXiv (the word list is provided in \cite{CTC-github}). Of the resulting papers returned in the search, if any of the tags are \textbf{cs.CR} (which broadly is defined as computer science regarding cryptography and cybersecurity), then we download that pdf and tag the document as cybersecurity related. 

For gathering non cybersecurity related documents from Arxiv, we searched all of the remaining non \textbf{cs.CR} categories and chose the top 100 (i.e., most relevant) papers from each of those categories; any of these papers with a \textbf{cs.CR} were not downloaded, since those documents would be cybersecurity related. 

\begin{figure}[h!]
    \centering
    \includegraphics[width=0.49\textwidth]{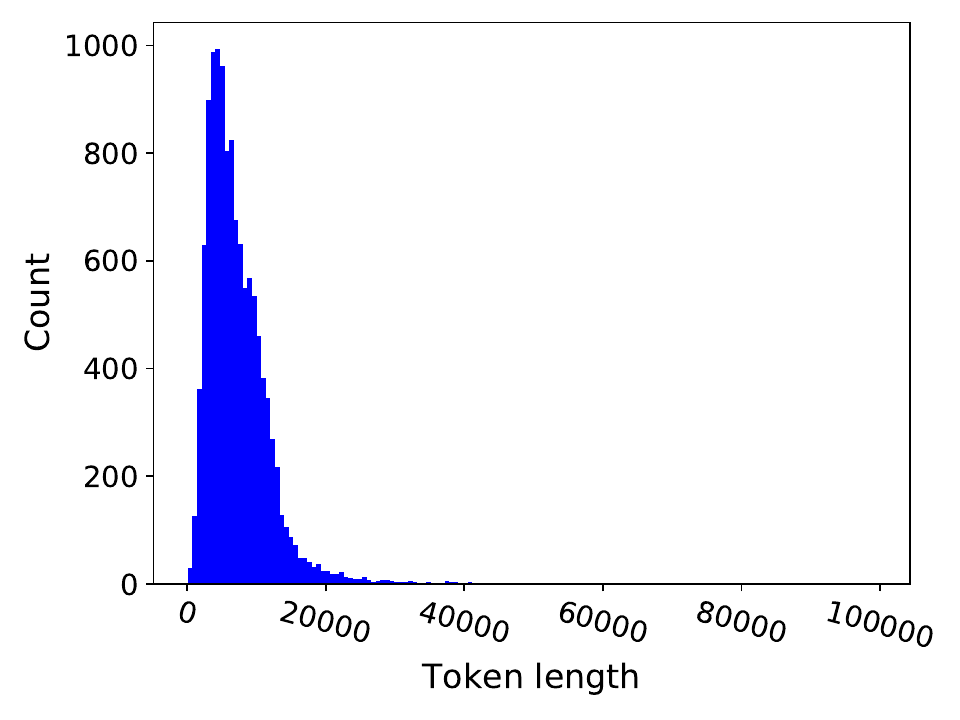}
    \includegraphics[width=0.49\textwidth]{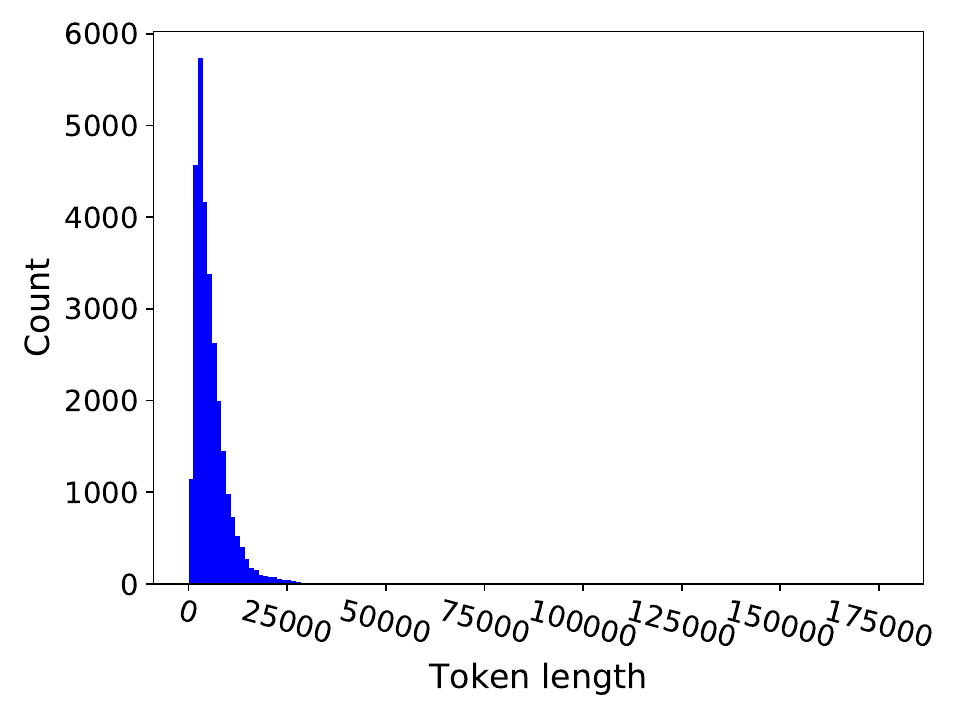}
    \caption{Arxiv token data histogram; cybersecurity documents (left) and non-cybersecurity documents (right). }
    \label{fig:Arxiv_token_length_histogram}
\end{figure}

The last step involves some text cleaning, which is specific to Arxiv, since all of the documents are PDFs. First, we remove all non-English documents (some of the downloaded technical documents were in a variety of other languages). Non-English documents were found using \textbf{langdetect} \cite{langdetect}. With non-English documents, the majority of the text was non-English, therefore the full document was not used. In future work, we may instead translate the non-English documents and use them as well. Next, we use \textbf{tika} to parse the PDF's into raw text. In some cases, \textbf{tika} is unable to parse the PDF's (in which case we can not use those documents). Figure \ref{fig:Arxiv_token_length_histogram} shows the distribution of the collected tagged arXiv text in terms of token (i.e., usable word) length. 

In Figure \ref{fig:Arxiv_token_length_histogram}, we see that the average document length for arXiv is in the thousands of words. In Figure \ref{fig:Reddit_token_length_histogram}, we saw that the Reddit average post length is less than 100 words, in contrast to Arxiv. In Figure \ref{fig:Stackexchange_token_length_histogram}, we observe that the average StackExchange post length is approximately 100. Across all three text sources, the difference in average document length between cybersecurity and non cybersecurity documents is marginal. The most significant difference between the cybersecurity and non cybersecurity documents is that there are many more non cybersecurity documents than cybersecurity documents. 

\subsection{Text preprocessing}
\label{sec:methods_text_proprocessing}
In order to use each document in the various classification algorithms, it is necessary to vectorize each document. In order to standardize all documents so that this vectorization process is consistent, we preprocess each document in a variety of ways. Specifically we remove URL's, non ascii characters, code tags, all HTML tags, and excessive whitespace. Table \ref{table:cleaned_text_sources} shows the total number of usable documents after we have cleaned the text.

\begin{table}[t]
\caption{Pre-processed text showing the number of cleaned and tokenized documents that are not empty (here by empty we mean having no detectable English words).}
\begin{center}
\begin{tabular}{ | l | l | l | }
\hline
Source & cybersecurity & non cybersecurity \\
\hline
\hline
Reddit & 163,739 & 4,129,605 \\ 
\hline
Stackexchange & 41,162 & 4,842,459 \\ 
\hline
Arxiv & 12,132 & 289,969 \\ 
\hline
\end{tabular}
\label{table:cleaned_text_sources}
\end{center}
\end{table}

\subsection{Vectorizer}
\label{sec:methods_vectorizer}
For this research, we used a term frequency - inverse document frequency (TF-IDF) vectorizer found in \textit{scikit-learn} \cite{scikit-learn}. TF-IDF vectorization is used to show how common (or important) a given word is in the input document. In particular, it weights words more heavily that occur at a greater frequency. The first step in creating a consistent vectorization method across all documents was compiling the more relevant and popular English words to use for this vectorizer. To this end, we used two English word lists and a custom list of cybersecurity terms. We used \textit{google20kwords} from \cite{english-dictionary2} and \textit{30k.txt} from \cite{english-dictionary1}. 

We merge all of these lists, remove all duplicates, and remove all entries that are not English words. To determine whether to remove or keep a word, if any of the three methods \textit{nltk} (see \cite{nltk}) words, nltk wordnet, or the python module \textit{enchant} consider a word to be English, we include it in the dictionary. Pronouns and nouns, such as names and company names, are not considered English words. The resulting English dictionary has a length of 24,538 (the word list is provided in \cite{CTC-github}). Non English words were not vectorized because the vast majority of the text we gathered was English and using this fixed English dictionary means that non English words are automatically removed. Future work may include handling foreign language words. 

We then fit a TF-IDF vectorizer to this dictionary, save it with 32 bit precision (in order to reduce memory costs), and use this vectorizer for computing the inputs to all machine learning classifiers. The TF-IDF vectorizer can be used to vectorize sentences, paragraphs, or entire documents. In our case, we vectorized each document using TF-IDF. We do not remove any stop words during this vectorization because in simple grid search experiments, removing stop words increased accuracy in some cases, while in other cases it decreased accuracy (both training and testing accuracy). Therefore, since removing stop words or not was not clearly motivated, we did not perform this extra step.

\subsection{Machine Learning Classifiers}
\label{sec:methods_ML_classifiers}
We test a total of 6 different types of machine learning classifiers. For all of these classifiers, we use a fixed TF-IDF vectorizer that has been fitted to the English dictionary described in the previous section. Specifically, we use \textit{Decision Tree}, \textit{Random Forest}, \textit{Logistic}, \textit{LinearSVC}, and \textit{Multi-layer Perceptron} (MLP). Each of these five models were provided in the python module \textbf{scikit-learn} (version 0.22.0) \cite{scikit-learn, sklearn_api}. We also build a Deep Neural Network (DNN) model using \textbf{tensorflow} (version 2.3.0) \cite{tensorflow2015-whitepaper}. 

For the Decision Tree classifier, we set the maximum depth to be 100, to reduce the real time computational cost. For the DNN model, we specify all of the parameters and hyperparameters used in Section \ref{sec:methods_DNN}. For all other models, we used default parameters. 

\subsubsection{Deep Neural Network}
\label{sec:methods_DNN}
We build a simple multi-layer neural network with one hidden layer using tensor flow; we use a sequential model with three layers. The first layer has an input equal to the length of English dictionary we used (24,538 words), with 10,000 nodes. The second layer has 1,000 nodes and the last layer has 100 nodes. The output layer has two nodes. The output layer uses an activation function \textit{softmax} and every other layer uses \textit{relu}. 

The number of epochs we use is not fixed when training the model. Instead, we train the model until a certain threshold training accuracy has been reached. In our experiments we try two different accuracy thresholds $0.95$ and $0.99$. The number of samples per iteration was fixed to $4,000$. We use \textit{sparse-categorical-crossentropy} for the model loss function, the model metric is \textit{accuracy}, and \textit{Adam} \cite{kingma2017adam} as the model optimizer. To speed up the training and validation time, we also set the flags \textit{workers} to $400$ and \textit{use-multiprocessing} to True. 

Since we try two different accuracy thresholds ($0.95$ and $0.99$), we actually train two different deep neural network models. Thus, in total we train seven different models in the following section. 
\section{Results}
\label{sec:experiments}

\begin{figure}[h!]
    \centering
    \includegraphics[width=0.49\textwidth]{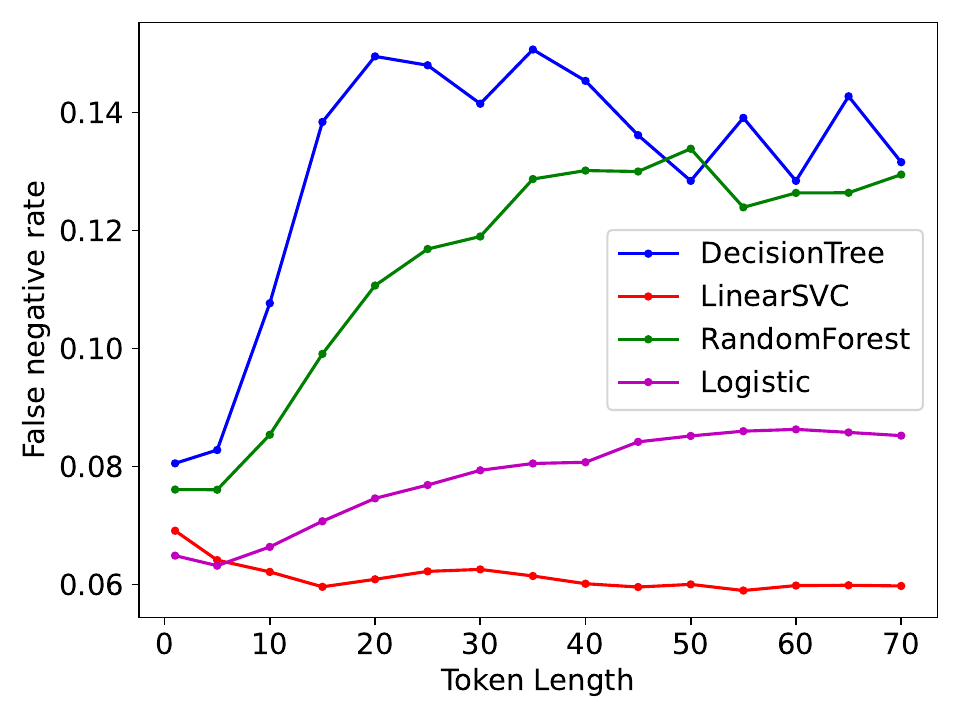}
    \includegraphics[width=0.49\textwidth]{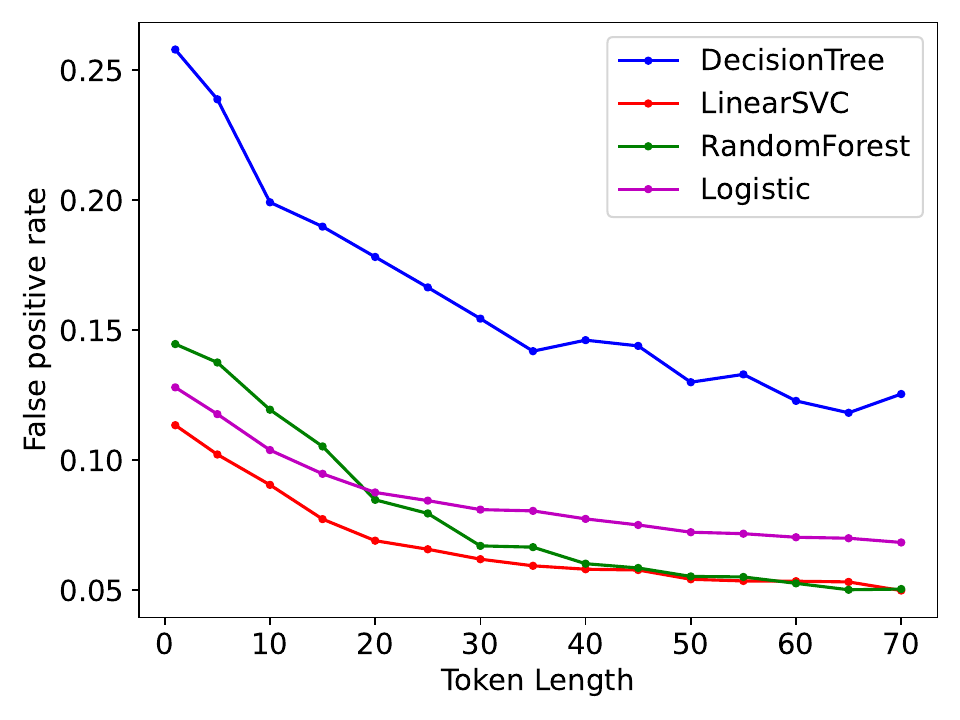}
    \caption{False negative rate as a function of token length (left) and False positive rate as a function of maximum token length (right). }
    \label{fig:Reddit_token_vs_acc}
\end{figure}

In this section we investigate some of the relevant parameters, both of the text and the machine learning models, that influence the training and testing accuracy.

First, we perform some simple grid-search optimizations of some of the model hyperparameters. Although not shown here for the sake of space, we found that these hyperparameters converge to reasonable performance; for example \textit{max-iters} should be set to a reasonably large number, but the training accuracy plateaus relatively quickly (i.e., after several hundred iterations). Second, in Section \ref{sec:experiments_token_length} we investigate how the minimum word length of the data changes both the training and the testing accuracy. In these first two steps, we generally had to run many instances of each algorithm and therefore typically we would select a subset of the data to train and validate on. In the third step, we train on half of each of the text source's datasets and validate the models on the other half (in some cases, due to RAM limitations, we trained on 3/8 of the data and validated on 5/8). Next, in Section \ref{sec:experiments_validation} we validate how each of these trained models performs on disparate text sources that they were not trained on (e.g., validating a model trained on Reddit text using text from StackExchange). In Section \ref{sec:experiments_confidence} we define a confidence measure from the continuous outputs of some of the machine learning models and determine how the confidence measure corresponds to error rates in the training dataset results. Finally, we combine all of these models into a unified Python tool called CTC. We show that CTC classifies large numbers of text documents with relatively low error rates. 

As shown in Table \ref{table:text_source_data}, the labelled data we have gathered is quite skewed towards being made of mostly not cybersecurity documents. This is not unexpected, but it means that training a machine learning model on that dataset will result in unequal weighting of the importance of the classification tasks. In an attempt to correct this towards an evenly balanced dataset, for each machine learning model we use the following class weighting rule. If we have $n_{c}$ cybersecurity documents and $n_{nc}$ not cybersecurity documents, then the class weight for not cybersecurity is $1$ and the class weight for cybersecurity is $\frac{n_{nc}}{c}$.

In the remainder of the article we use FN to denote false negative rate and FP to denote false positive rate. 

\subsection{Token Length}
\label{sec:experiments_token_length}

\begin{figure}[h!]
    \centering
    \includegraphics[width=0.49\textwidth]{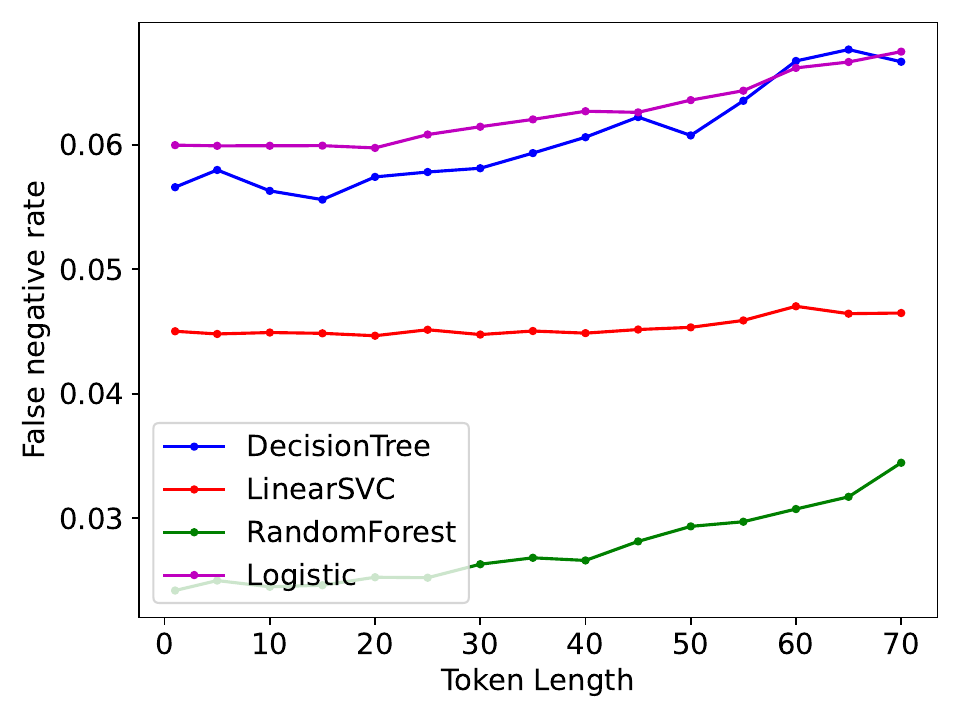}
    \includegraphics[width=0.49\textwidth]{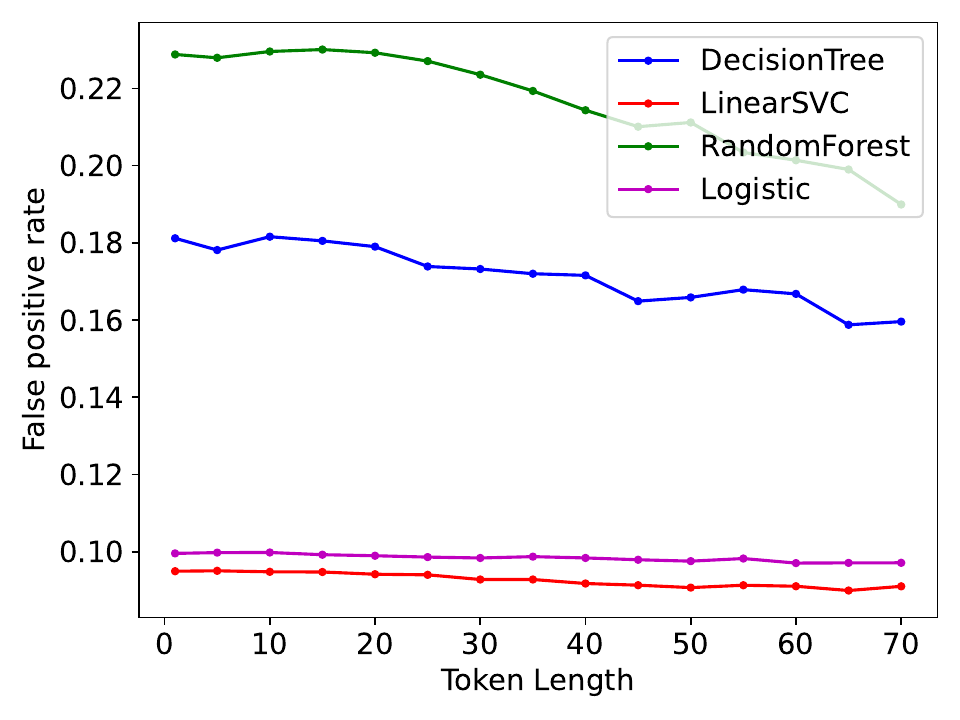}
    \caption{False negative (left) and false positive (right) rates as a function of maximum token length for the StackExchange labelled text}
    \label{fig:Stackexchange_token_vs_acc}
\end{figure}

\begin{figure}[h!]
    \centering
    \includegraphics[width=0.49\textwidth]{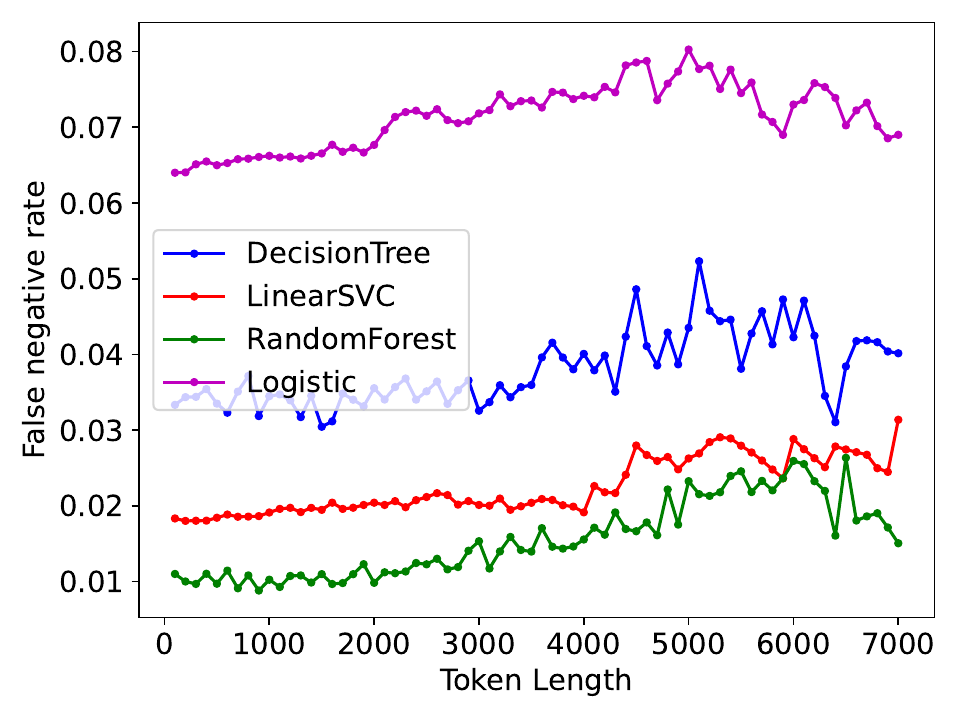}
    \includegraphics[width=0.49\textwidth]{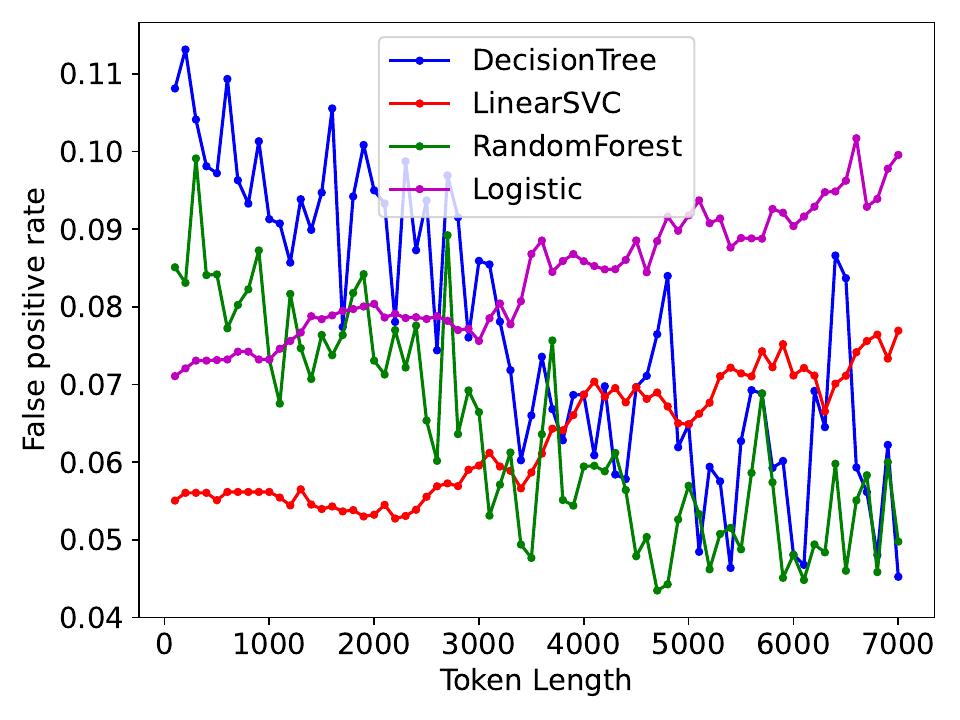}
    \caption{False negative (left) and false positive (right) rates as a function of maximum token length for the arXiv labelled text}
    \label{fig:Arxiv_token_vs_acc}
\end{figure}

In this section, we determine the behavior of several of the classifiers as a function of the minimum token length used on the training and validation datasets. Across all three text sources we select random subsets of the full training data to reduce the total needed computation time. The procedure we used was to eliminate all documents with usable token lengths less than $N$ ($N$ is plotted on the x-axis of the plots shown in the remaining subsections), train the model on the training data, and then validate (and plot) the accuracy using the unseen validation data (both the validation data and training data had at least $N$ usable tokens in each document). To save space, we show these results for four different machine learning classifiers, \textit{Decision Tree}, \textit{LinearSVC}, \textit{Random Forest}, \textit{Logistic}, on each of three text sources. 

\subsubsection{Reddit}
For Reddit, we set both the training and validation datasets to have $40,000$ Cybersecurity labelled documents and $50,000$ non Cybersecurity labelled documents. In Figure \ref{fig:Reddit_token_vs_acc}, we plot the false negative and false positive rate as a function of token length for four different classifiers. Figure \ref{fig:Reddit_token_vs_acc} shows on the left an increase in the false negative rate for the Random Forest Decision Tree and Logistic classifiers as token length increases, while LinearSVC's false negative rate remains relatively unchanged. Figure \ref{fig:Reddit_token_vs_acc} shows on the right a decrease in the false positive rate as a function of token length across all four classifiers.

\subsubsection{StackExchange}
For StackExchange, we set both the training and validation datasets to have $20,581$ Cybersecurity labelled documents and $50,000$ non Cybersecurity labelled documents. In Figure \ref{fig:Stackexchange_token_vs_acc}, we plot the false negative and false positive rate as a function of token length for four different classifiers. Figure \ref{fig:Stackexchange_token_vs_acc} shows on the left that Random Forest, Logistic, and Decision Tree models increase in false negative rate as token length increases. Figure \ref{fig:Stackexchange_token_vs_acc} shows on the right that the false positive rate decreases as a function of token length for DecisionTree and RandomForest, while LinearSVC and Logistic classifiers remain relatively unchanged. 

\subsubsection{Arxiv}
For Arxiv, we set both the training and validation datasets to have $1,000$ cybersecurity labelled documents, and $3,000$ not cybersecurity labelled documents. In Figure \ref{fig:Stackexchange_token_vs_acc} we plot the false negative and false positive rate as a function of token length for four different classifiers. Figure \ref{fig:Stackexchange_token_vs_acc} shows on the left an increase in false negative rates as a function of usable token length for each of the four classifiers. Figure \ref{fig:Stackexchange_token_vs_acc} shows on the right that DecisionTree and RandomForest false positive rates decrease as a function of token length, while LinearSVC and Logistic classifiers increase in false positive rates as a function of token length. Since the average usable word length of the documents from arXiv are very large, see Figure \ref{fig:Arxiv_token_length_histogram}, the manner in which the error rates change as a function of token length for the arXiv source is not as important as for Reddit and StackExchange sources. 

\begin{table*}[t!]
\centering
\caption{Cross validate DNN with training accuracy $0.99$}
\label{table:cross_validate_DNN_0.99}
\begin{tabular}{|l|l|l|l|l|}
    \hline
    Training source & Validation source & Training accuracy & Validation FN & Validation FP \\
    \hline
    \hline
    Reddit 1/2 & arXiv &  & 0.0505 & 0.1438 \\
    \hline
    Reddit 1/2 & StackExchange &  & 0.2903 & 0.1115 \\
    \hline
    Reddit 1/2 & Reddit 1/2 & 0.9932 & 0.0177 & 0.1801 \\
    \hline
    \hline
    StackExchange 1/2 & arXiv &  & 0.0007 & 0.7711 \\
    \hline
    StackExchange 1/2 & StackExchange 1/2 & 0.991 & 0.0083 & 0.3081 \\
    \hline
    StackExchange 1/2 & Reddit &  & 0.0666 & 0.6837 \\
    \hline
    \hline
    arXiv 1/2 & arXiv 1/2 & 0.9933 & 0.0172 & 0.0236 \\
    \hline
    arXiv 1/2 & StackExchange &  & 0.2353 & 0.2007 \\
    \hline
    arXiv 1/2 & Reddit &  & 0.1068 & 0.4041 \\
\hline
\end{tabular}
\end{table*}

\begin{table*}[t!]
\centering
\caption{Cross validate DNN with training accuracy $0.95$}
\label{table:cross_validate_DNN_0.95}
\begin{tabular}{|l|l|l|l|l|}
    \hline
    Training source & Validation source & Training accuracy & Validation FN & Validation FP \\
    \hline
    \hline
    Reddit 1/2 & arXiv &  & 0.0867 & 0.0886 \\
    \hline
    Reddit 1/2 & StackExchange &  & 0.4485 & 0.0499 \\
    \hline
    Reddit 1/2 & Reddit 1/2 & 0.9573 & 0.0397 & 0.1043 \\
    \hline
    \hline
    StackExchange 1/2 & arXiv &  & 0.0013 & 0.6222 \\
    \hline
    StackExchange 1/2 & StackExchange 1/2 & 0.9644 & 0.0317 & 0.1415 \\
    \hline
    StackExchange 1/2 & Reddit &  & 0.0662 & 0.5342 \\
    \hline
    \hline
    arXiv 1/2 & arXiv 1/2 & 0.9578 & 0.0182 & 0.0748 \\
    \hline
    arXiv 1/2 & StackExchange &  & 0.2052 & 0.2393 \\
    \hline
    arXiv 1/2 & Reddit &  & 0.0714 & 0.4964 \\
\hline
\end{tabular}
\end{table*}

To train the machine learning models, we want to not include documents with too few tokens because the topic being discussed may be ambiguous or unclear. However, we want relatively balanced error rates between false negative and false positive. Most importantly, we want our models to be able to classify a wide range of internet text. For this reason, we would want to train on data with smaller token lengths. Given these factors, we conclude that a minimum usable token length of ten is reasonable for training the models in the remaining experiments.

\subsection{Validation}
\label{sec:experiments_validation}

Next, we train all seven of the machine learning models described in Section \ref{sec:methods_ML_classifiers} on each of the three labelled text sources (Reddit, StackExchange, Arxiv), resulting in a total of 21 distinct trained models. Following the token length experiments of Section \ref{sec:experiments_token_length}, we use only labelled data which has word length (token length) of at least ten (this reduces the total amount of labelled data we use from the initial corpus shown in Table \ref{table:cleaned_text_sources}). For each model type and each text source, we train the model on one half of the labelled data, and then validate on the other half (in some cases, where training on one half of the data is too computationally costly, we train on 3/8 of the data).

Since we want these machine learning models to be applied to text that may not come form arXiv or StackExchange or Reddit, one way to evaluate each of these models in a more robust way is to predict the topic of text from the two sources the model was not trained on; for example predicting the topic of the labelled arXiv and StackExchange dataset given the model was trained on Reddit text. In this section, we show the validation accuracy (i.e., testing accuracy) and cross validation accuracy for each of the 21 trained models. In particular, Tables \ref{table:cross_validate_DNN_0.99}, \ref{table:cross_validate_DNN_0.95}, \ref{table:cross_validate_Logistic}, \ref{table:cross_validate_RandomForest}, \ref{table:cross_validate_LinearSVC}, \ref{table:cross_validate_DecisionTree}, \ref{table:cross_validate_MLP} show the cross validation results. In these tables we show the training accuracy as well as the cross validation false negative and false positive rates. Under the training accuracy column, we only get one training accuracy entry for each of the three text sources (the other rows are validation accuracy results), which means that six of the entries in that column will be empty. 

\begin{table*}[t]
\centering
\caption{Cross validate Logistic}
\label{table:cross_validate_Logistic}
\begin{tabular}{|l|l|l|l|l|}
    \hline
    Training source & Validation source & Training accuracy & Validation FN & Validation FP \\
    \hline
    \hline
    Reddit 1/2 & arXiv &  & 0.27 & 0.0101 \\
    \hline
    Reddit 1/2 & StackExchange &  & 0.5136 & 0.0282 \\
    \hline
    Reddit 1/2 & Reddit 1/2 & 0.9509 & 0.0496 & 0.0849 \\
    \hline
    \hline
    StackExchange 3/8 & arXiv &  & 0.0255 & 0.2211 \\
    \hline
    StackExchange 3/8 & StackExchange 5/8 & 0.9599 & 0.0403 & 0.1037 \\
    \hline
    StackExchange 3/8 & Reddit &  & 0.1162 & 0.4089 \\
    \hline
    \hline
    arXiv 1/2 & arXiv 1/2 & 0.965 & 0.0296 & 0.0557 \\
    \hline
    arXiv 1/2 & StackExchange &  & 0.0347 & 0.6919 \\
    \hline
    arXiv 1/2 & Reddit &  & 0.0035 & 0.8513 \\
\hline
\end{tabular}
\end{table*}

\begin{table*}[t]
\centering
\caption{Cross validate RandomForest}
\label{table:cross_validate_RandomForest}
\begin{tabular}{|l|l|l|l|l|}
    \hline
    Training source & Validation source & Training accuracy & Validation FN & Validation FP \\
    \hline
    \hline
    Reddit 1/2 & arXiv &  & 0.0037 & 0.7019 \\
    \hline
    Reddit 1/2 & StackExchange &  & 0.0271 & 0.7481 \\
    \hline
    Reddit 1/2 & Reddit 1/2 & 0.9998 & 0.0018 & 0.6921 \\
    \hline
    \hline
    StackExchange 1/2 & arXiv &  & 0.0001 & 0.9736 \\
    \hline
    StackExchange 1/2 & StackExchange 1/2 & 1.0 & 0 & 0.9811 \\
    \hline
    StackExchange 1/2 & Reddit &  & 0.0004 & 0.982 \\
    \hline
    \hline
    arXiv 1/2 & arXiv 1/2 & 1.0 & 0.013 & 0.0374 \\
    \hline
    arXiv 1/2 & StackExchange &  & 0.001 & 0.9705 \\
    \hline
    arXiv 1/2 & Reddit &  & 0.0001 & 0.9914 \\
\hline
\end{tabular}
\end{table*}

As a general summary of the cross validation results, we observe the trend that the false negative and false positive rates for the models trained on a source and then validated on the same text source (not the same data, just the same text source e.g., Reddit) are usually relatively low (less than ten percent). The exceptions to this are RandomForest and Decision Tree for Reddit and StackExchange sources. We also observe varied error rates for the cross validation experiments. In particular, we usually see an asymmetry in the error rates - i.e., the false negative rate is very high and the false positive rate is very low, or the reverse. 

Table \ref{table:cross_validate_DNN_0.95} and \ref{table:cross_validate_DNN_0.99} show the direct comparison of validation and training accuracy differences when applying the $0.95$ and $0.99$ training accuracy cutoff. The most notable differences between these two models is that the $0.95$ training accuracy model has less overall false positive error rate when predicting on the Reddit and StackExchange text, whereas the arXiv testing false positive rates were worse. For both the $0.99$ and $0.95$ trained models we see that the worse performing regions are the StackExchange trained models predicting on Reddit and arXiv text, which gives a false positive rate greater than $0.50$ in all cases. The false positive rates for the arXiv text trained models predicting on Reddit text also gives very high false positive rates (greater than $0.40$). 

Tables \ref{table:cross_validate_LinearSVC}, \ref{table:cross_validate_Logistic}, \ref{table:cross_validate_MLP}, and \ref{table:cross_validate_RandomForest} all follow similar trends to the DNN models in terms of the highest error rate validation results. In particular, the StackExchange trained models predicting on Reddit and arXiv text, and the arXiv trained models predicting on Reddit text all give very high false positive rates. The false negative error rates on the other hand are significantly lower than the false positive  rates for most of the models with a few exceptions such as LinearSVC model trained on Reddit text predicting on StackExchange text (see Table \ref{table:cross_validate_LinearSVC}). 

\begin{table*}[t]
\centering
\caption{Cross validate LinearSVC}
\label{table:cross_validate_LinearSVC}
\begin{tabular}{|l|l|l|l|l|}
    \hline
    Training source & Validation source & Training accuracy & Validation FN & Validation FP \\
    \hline
    \hline
    Reddit 1/2 & arXiv &  & 0.1482 & 0.0463 \\
    \hline
    Reddit 1/2 & StackExchange &  & 0.4543 & 0.0397 \\
    \hline
    Reddit 1/2 & Reddit 1/2 & 0.9566 & 0.0463 & 0.0992 \\
    \hline
    \hline
    StackExchange 3/8 & arXiv &  & 0.0041 & 0.5392 \\
    \hline
    StackExchange 3/8 & StackExchange 5/8 & 0.9673 & 0.0335 & 0.1385 \\
    \hline
    StackExchange 3/8 & Reddit &  & 0.0928 & 0.4993 \\
    \hline
    \hline
    arXiv 1/2 & arXiv 1/2 & 0.9882 & 0.0147 & 0.0302 \\
    \hline
    arXiv 1/2 & StackExchange &  & 0.0205 & 0.6619 \\
    \hline
    arXiv 1/2 & Reddit &  & 0.0039 & 0.8224 \\
\hline
\end{tabular}
\end{table*}

\begin{table*}[t]
\centering
\caption{Cross validate DecisionTree}
\label{table:cross_validate_DecisionTree}
\begin{tabular}{|l|l|l|l|l|}
    \hline
    Training source & Validation source & Training accuracy & Validation FN & Validation FP \\
    \hline
    \hline
    Reddit 1/2 & arXiv &  & 0.2338 & 0.2309 \\
    \hline
    Reddit 1/2 & StackExchange &  & 0.4181 & 0.3088 \\
    \hline
    Reddit 1/2 & Reddit 1/2 & 0.9749 & 0.0398 & 0.3233 \\
    \hline
    \hline
    StackExchange 1/2 & arXiv &  & 0.0242 & 0.2986 \\
    \hline
    StackExchange 1/2 & StackExchange 1/2 & 0.9854 & 0.0188 & 0.4127 \\
    \hline
    StackExchange 1/2 & Reddit &  & 0.0248 & 0.6518 \\
    \hline
    \hline
    arXiv 1/2 & arXiv 1/2 & 1.0 & 0.0249 & 0.0641 \\
    \hline
    arXiv 1/2 & StackExchange &  & 0.0186 & 0.7244 \\
    \hline
    arXiv 1/2 & Reddit &  & 0.0132 & 0.7857 \\
\hline
\end{tabular}
\end{table*}

\begin{table*}[t!]
\centering
\caption{Cross validate Multi-Layer Perceptron}
\label{table:cross_validate_MLP}
\begin{tabular}{|l|l|l|l|l|}
    \hline
    Training source & Validation source & Training accuracy & Validation FN & Validation FP \\
    \hline
    \hline
    Reddit 1/2 & arXiv &  & 0.0343 & 0.2406 \\
    \hline
    Reddit 1/2 & StackExchange &  & 0.1412 & 0.276 \\
    \hline
    Reddit 1/2 & Reddit 1/2 & 0.9996 & 0.0078 & 0.3183 \\
    \hline
    \hline
    StackExchange 3/8 & arXiv &  & 0.0004 & 0.8492 \\
    \hline
    StackExchange 3/8 & StackExchange 5/8 & 0.9995 & 0.0027 & 0.5325 \\
    \hline
    StackExchange 3/8 & Reddit &  & 0.0179 & 0.8125 \\
    \hline
    \hline
    arXiv 1/2 & arXiv 1/2 & 0.9999 & 0.0147 & 0.0432 \\
    \hline
    arXiv 1/2 & StackExchange &  & 0.3002 & 0.3592 \\
    \hline
    arXiv 1/2 & Reddit &  & 0.117 & 0.5524 \\
\hline
\end{tabular}
\end{table*}

\subsection{Classification Confidence Measure}
\label{sec:experiments_confidence}

\begin{figure}[t!]
    \centering
    \includegraphics[width=0.49\textwidth]{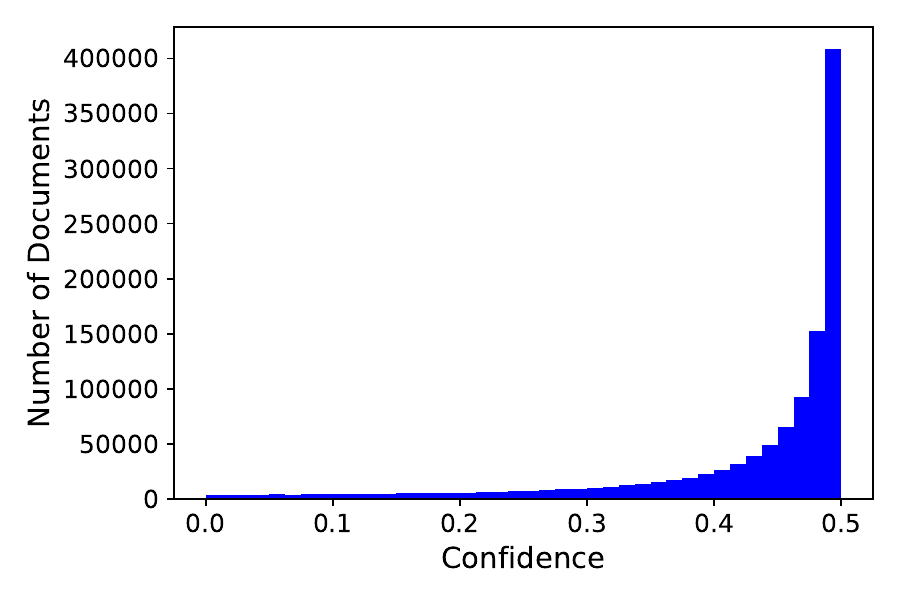}
    \includegraphics[width=0.49\textwidth]{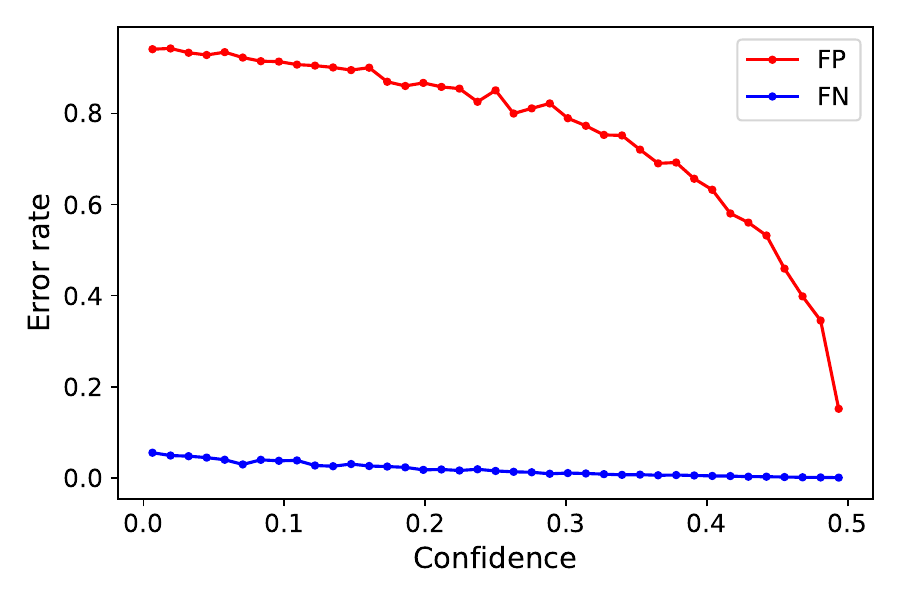}\\
    \includegraphics[width=0.49\textwidth]{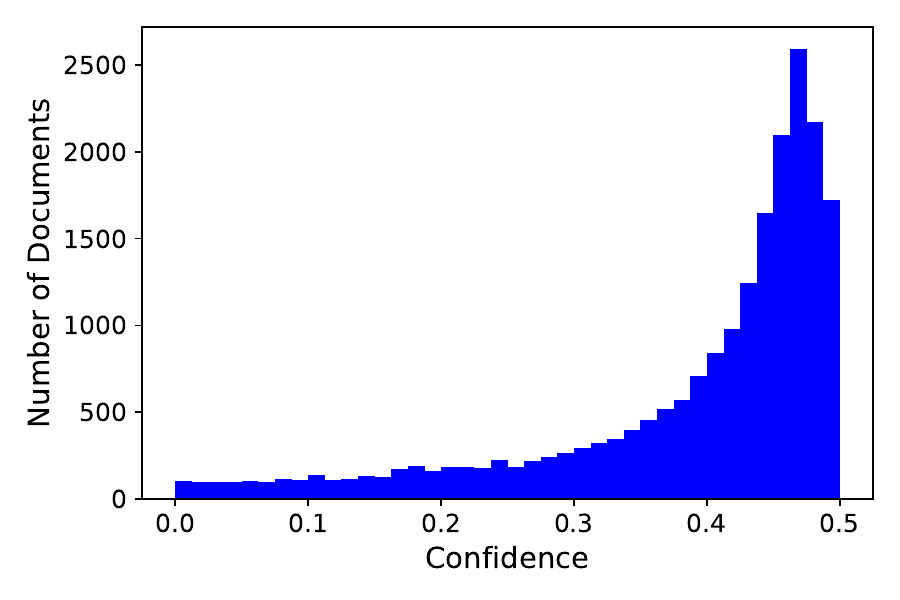}
    \includegraphics[width=0.49\textwidth]{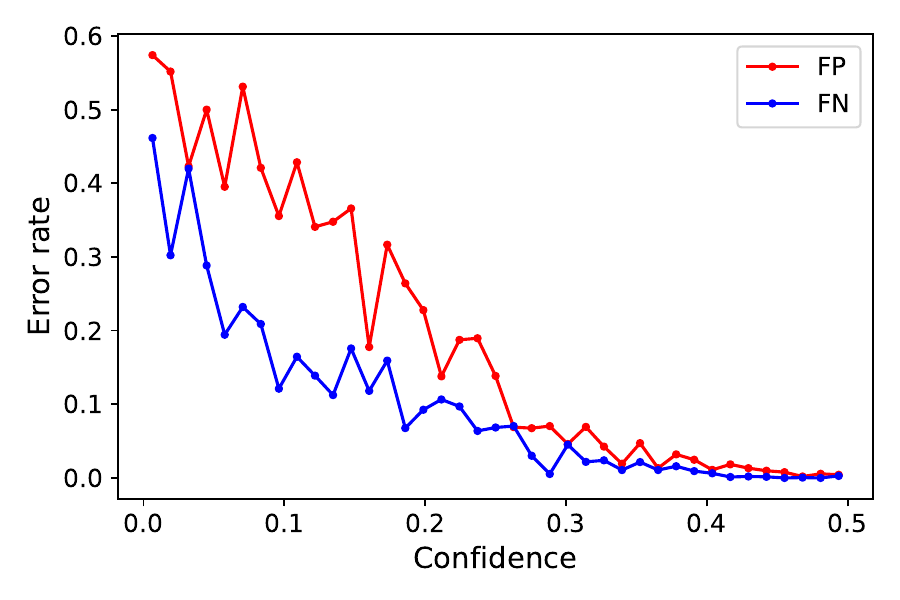}\\
    \includegraphics[width=0.49\textwidth]{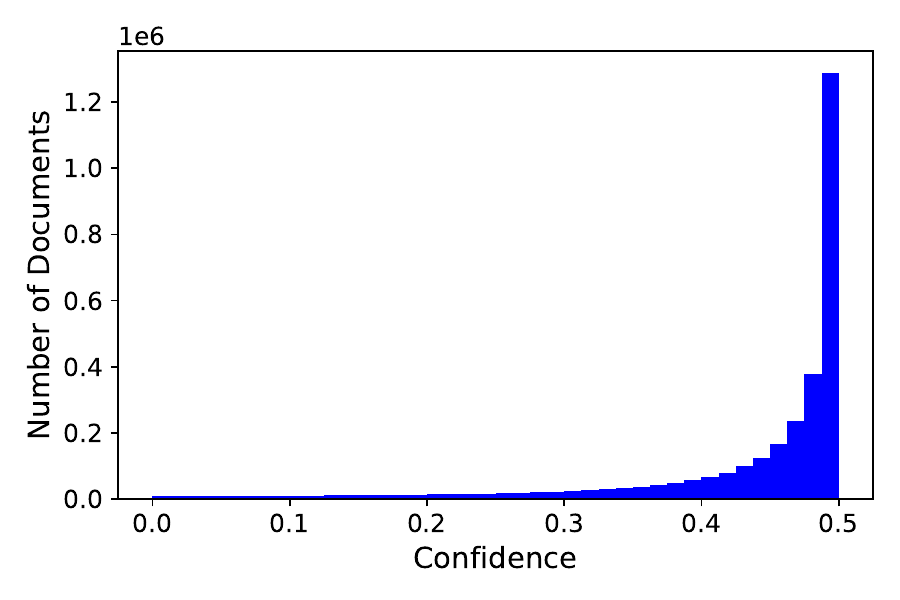}
    \includegraphics[width=0.49\textwidth]{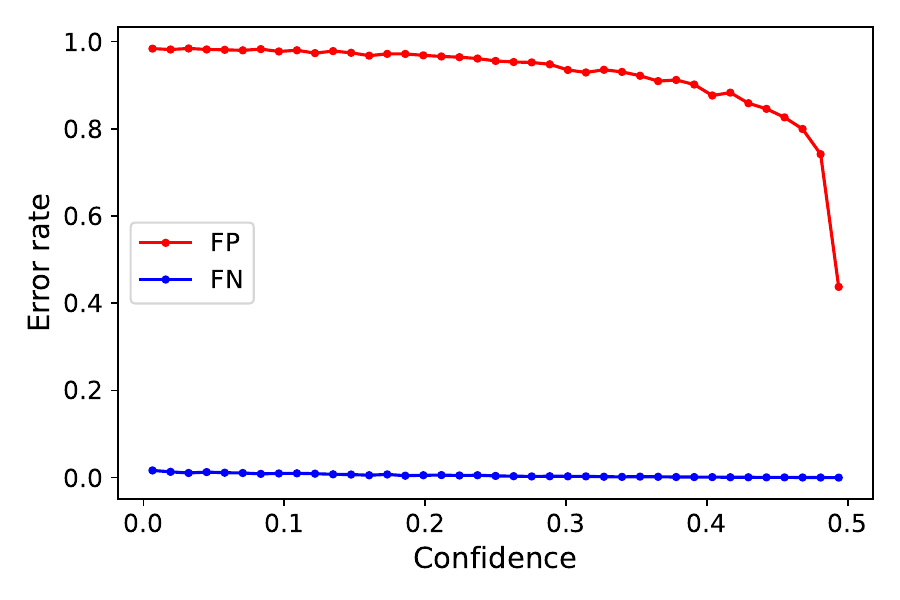}\\
    \caption{Histogram of confidence measures (left column) and error rates for each bin of confidence measures (right column). Logistic trained and then validated on Reddit text (top row), arXiv text (middle row), and StackExchange text (bottom row).}
    \label{fig:confidence_validation_Logistic}
\end{figure}

\begin{figure}[t]
    \centering
    \includegraphics[width=0.49\textwidth]{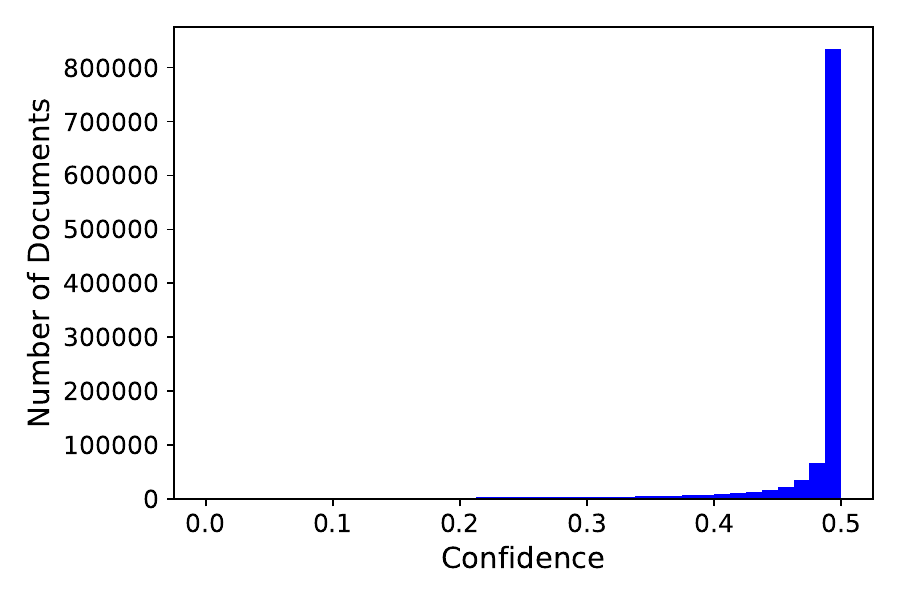}
    \includegraphics[width=0.49\textwidth]{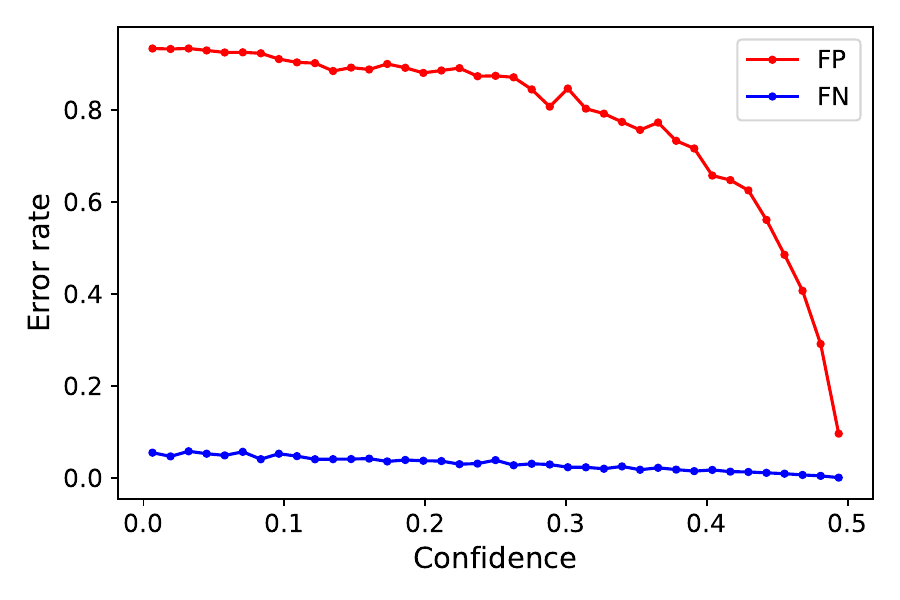}\\
    \includegraphics[width=0.49\textwidth]{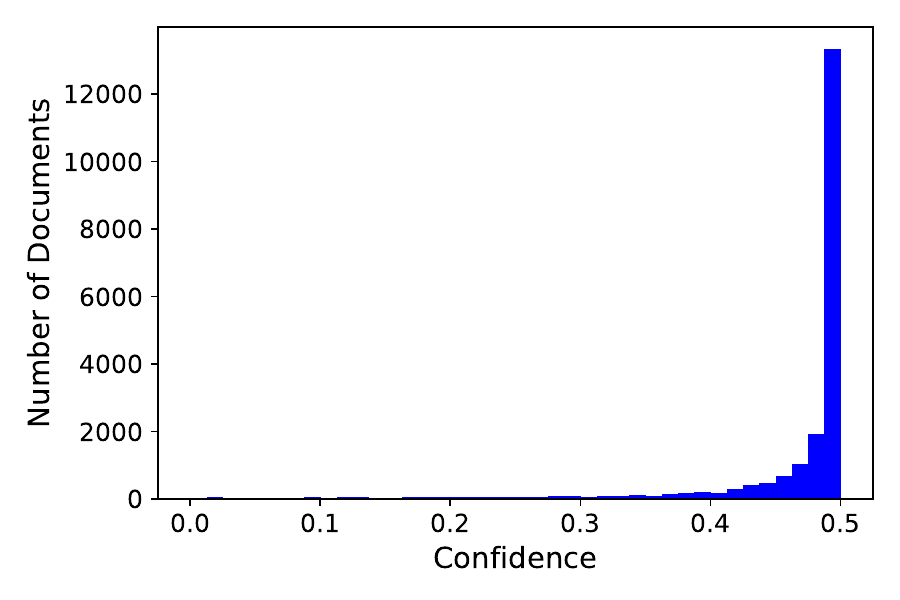}
    \includegraphics[width=0.49\textwidth]{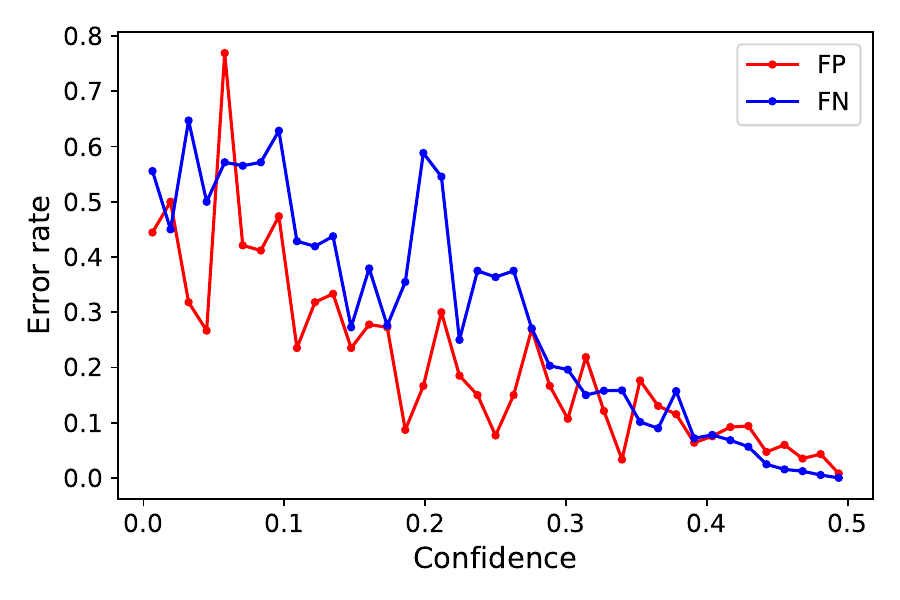}\\
    \includegraphics[width=0.49\textwidth]{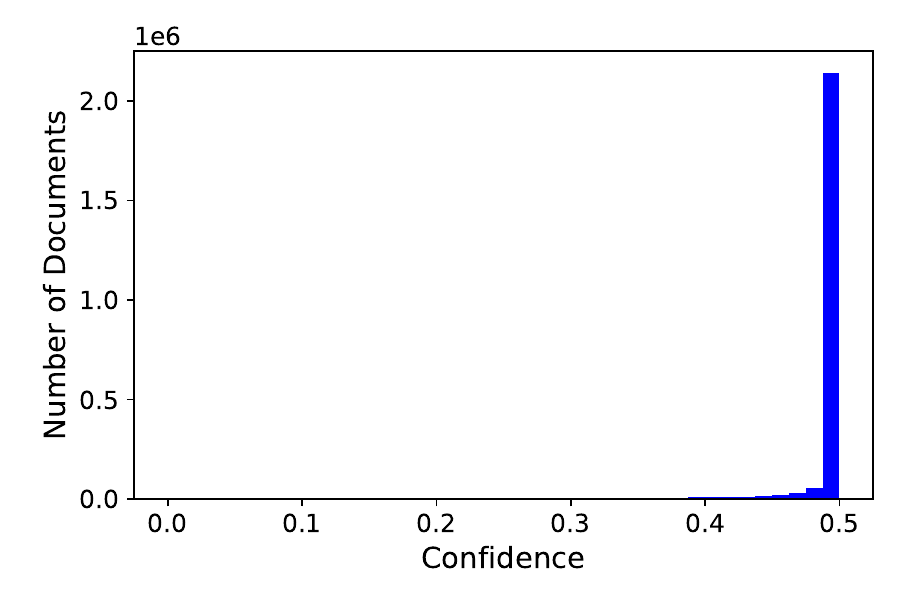}
    \includegraphics[width=0.49\textwidth]{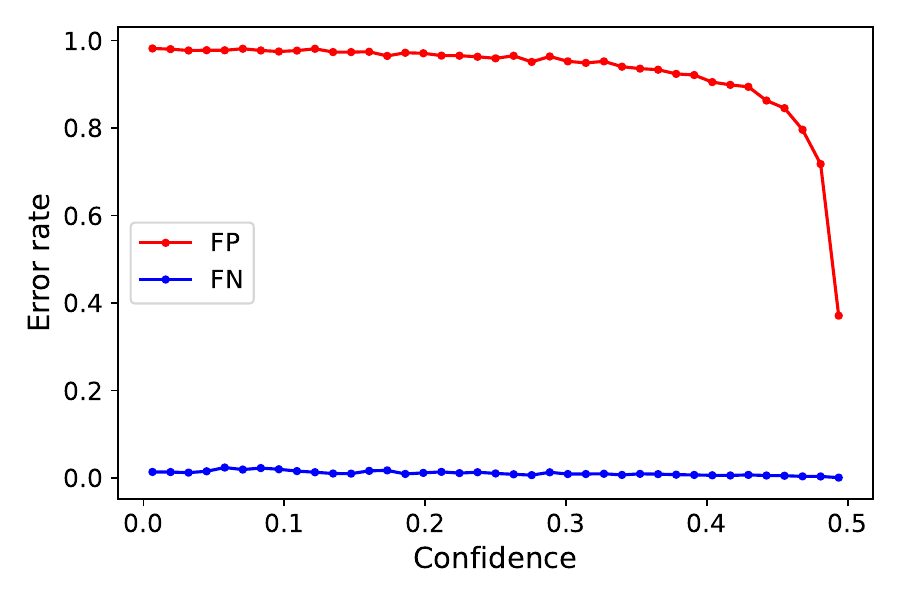}\\
    \caption{Histogram of confidence measures (left column) and error rates for each bin of confidence measures (right column). DNN model trained  to $0.95$ training accuracy and then validated on Reddit text (top row), arXiv text (middle row), and StackExchange text (bottom row).}
    \label{fig:confidence_validation_DNN0.95}
\end{figure}

\begin{figure}[t]
    \centering
    \includegraphics[width=0.49\textwidth]{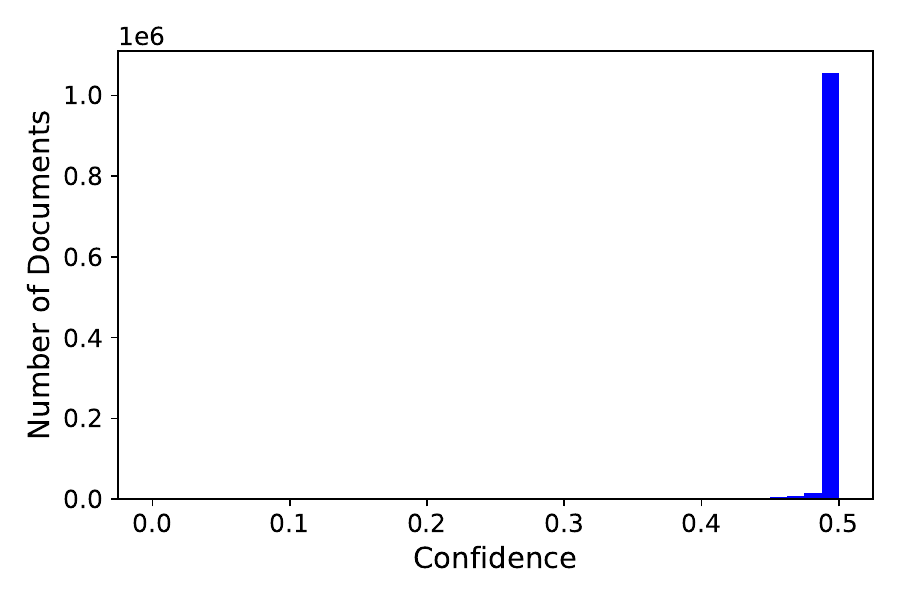}
    \includegraphics[width=0.49\textwidth]{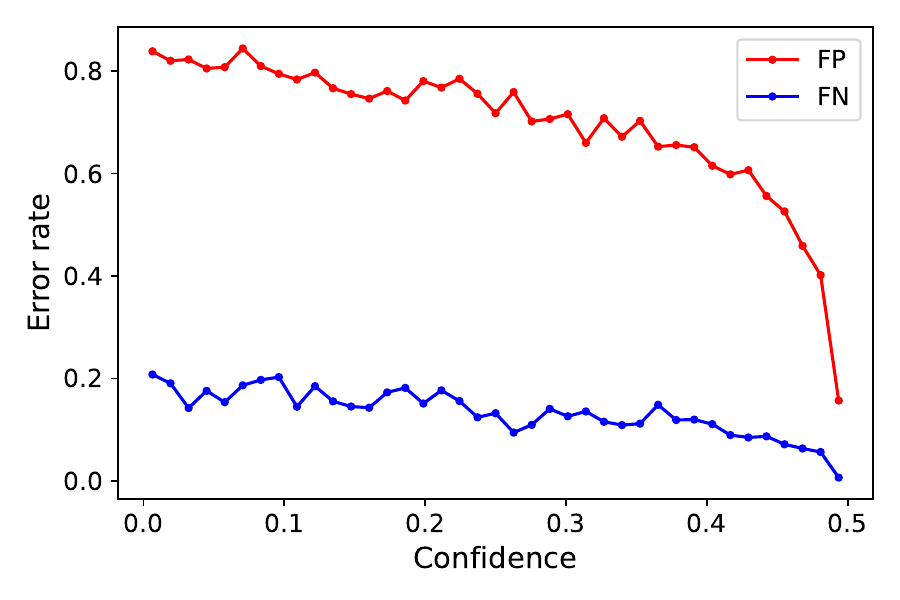}\\
    \includegraphics[width=0.49\textwidth]{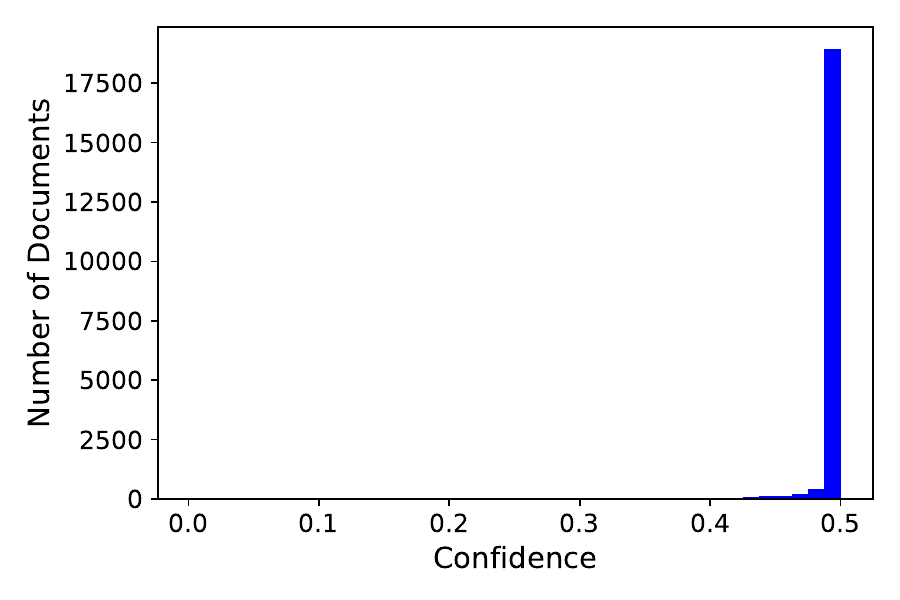}
    \includegraphics[width=0.49\textwidth]{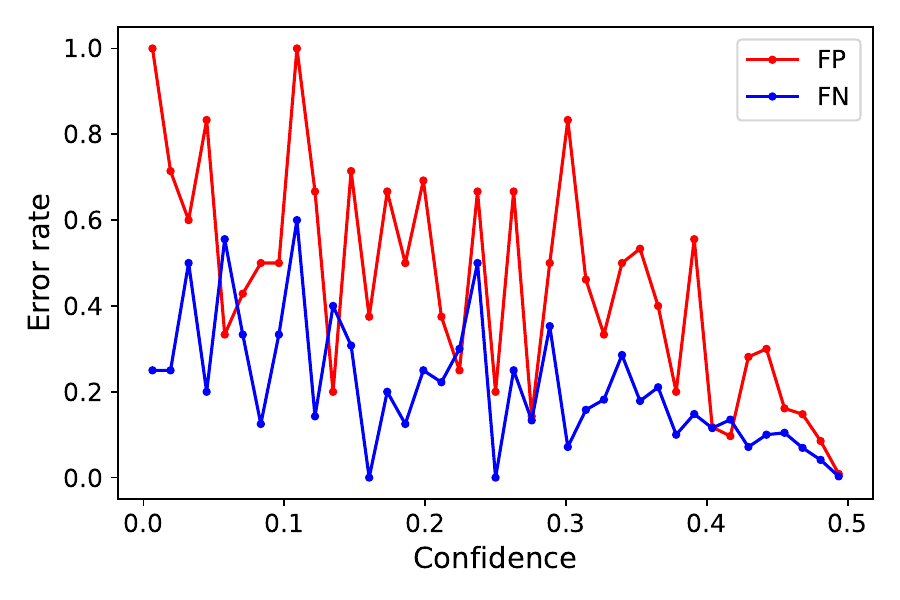}\\
    \includegraphics[width=0.49\textwidth]{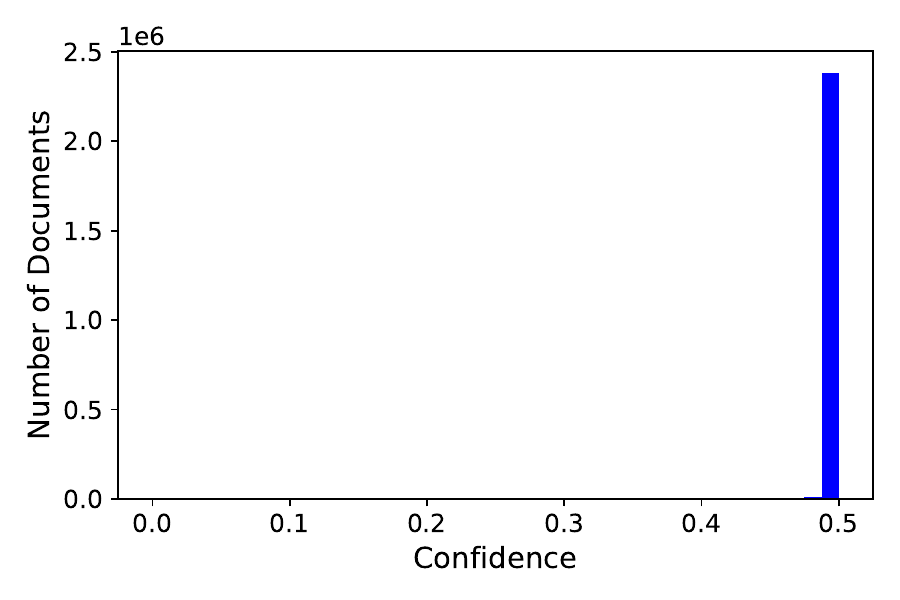}
    \includegraphics[width=0.49\textwidth]{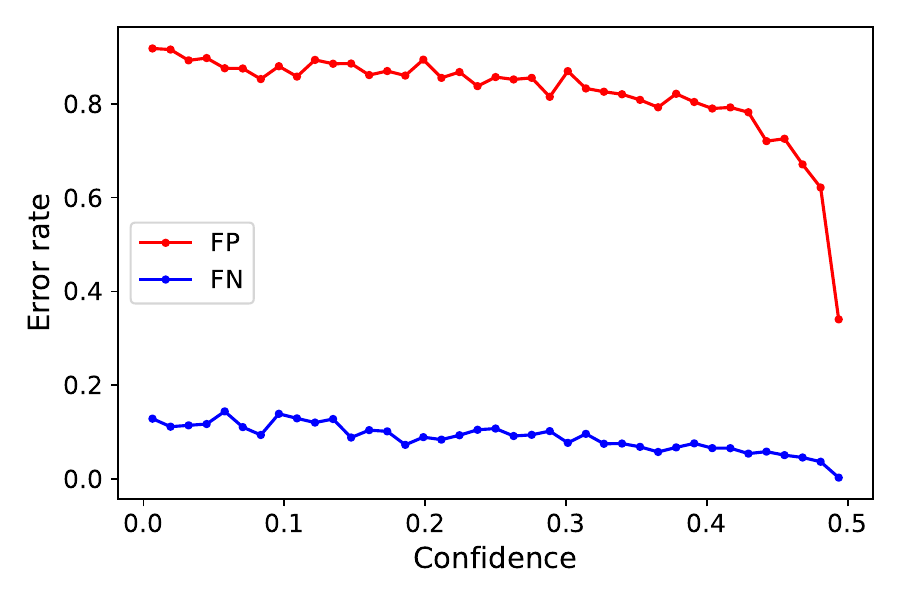}\\
    \caption{Histogram of confidence measures (left column) and error rates for each bin of confidence measures (right column). DNN model trained  to $0.99$ training accuracy and then validated on Reddit text (top row), arXiv text (middle row), and StackExchange text (bottom row).}
    \label{fig:confidence_validation_DNN0.99}
\end{figure}

\begin{figure}[t]
    \centering
    \includegraphics[width=0.49\textwidth]{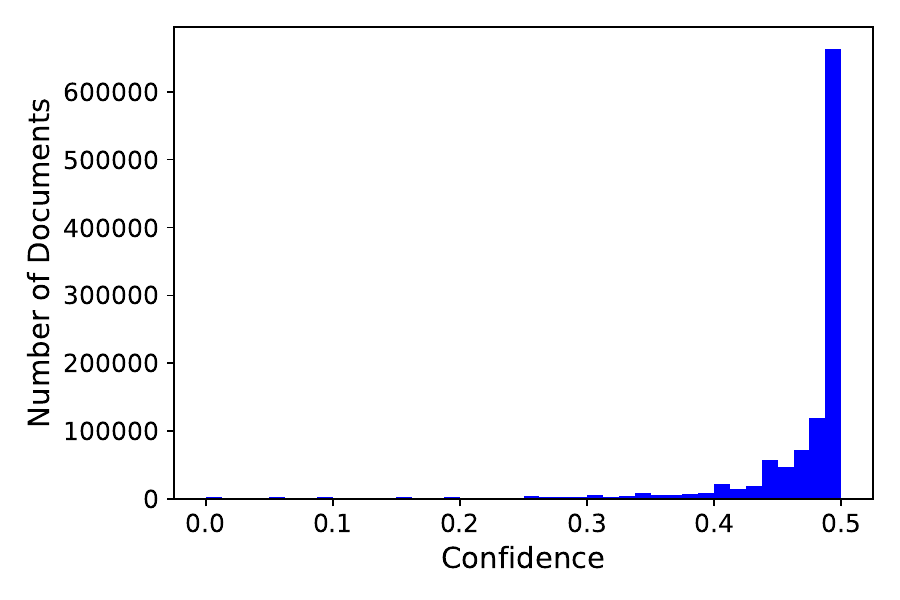}
    \includegraphics[width=0.49\textwidth]{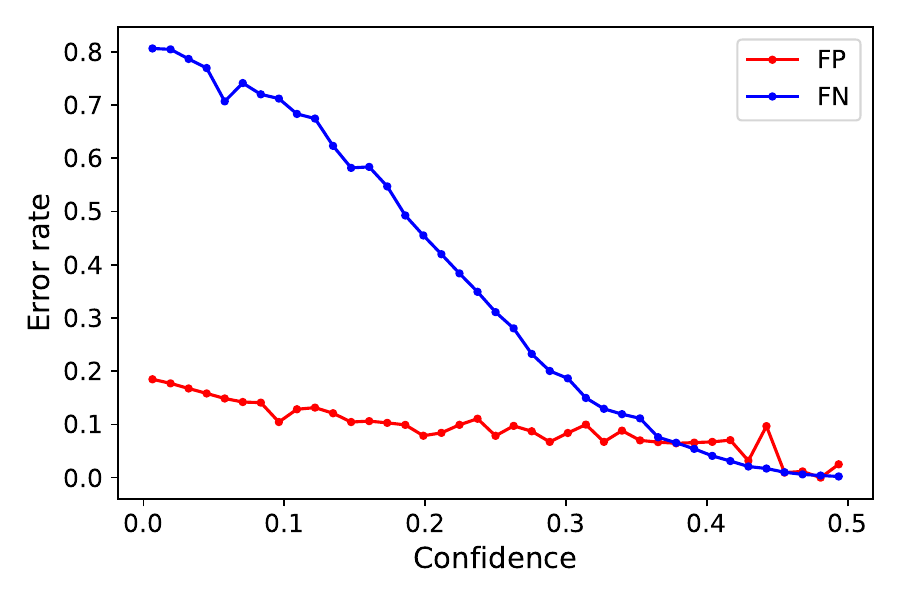}\\
    \includegraphics[width=0.49\textwidth]{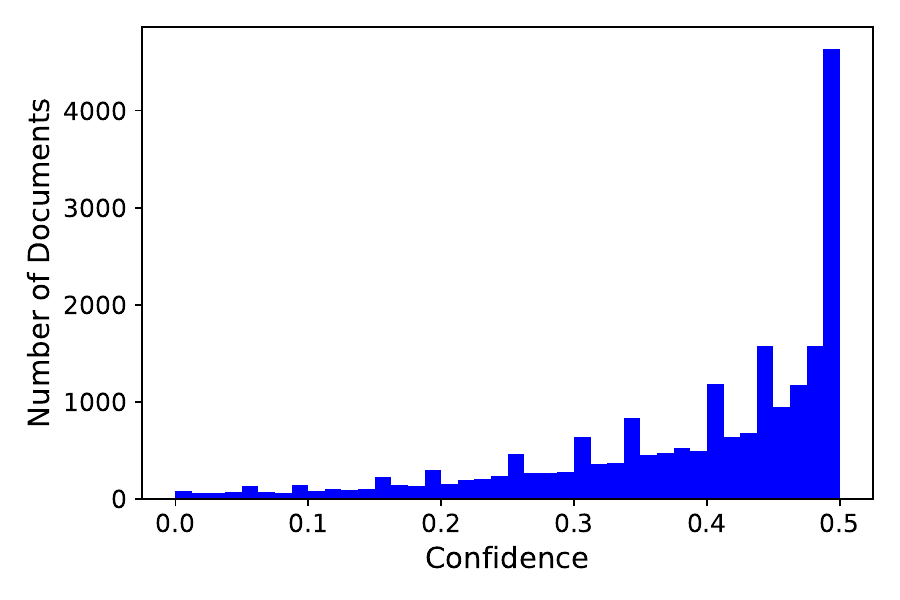}
    \includegraphics[width=0.49\textwidth]{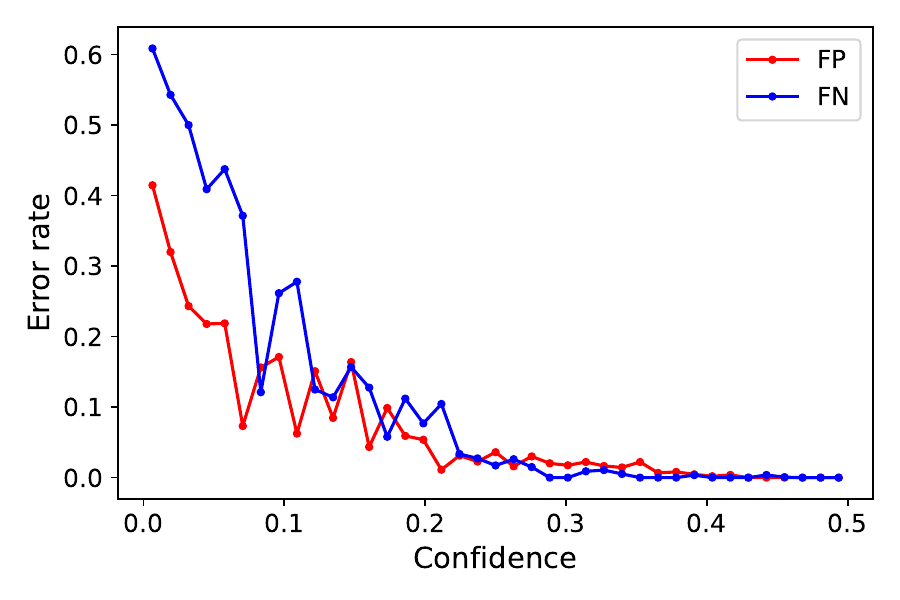}\\
    \includegraphics[width=0.49\textwidth]{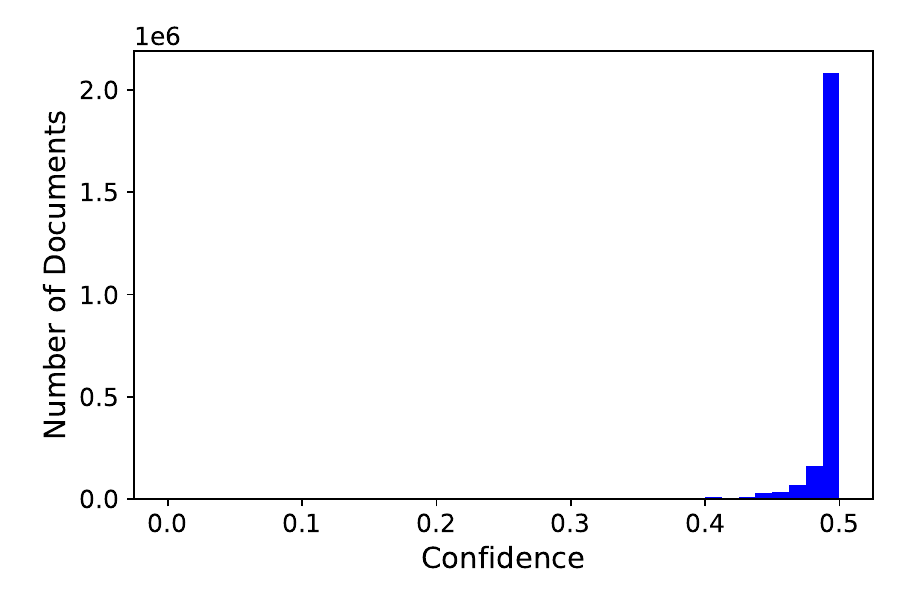}
    \includegraphics[width=0.49\textwidth]{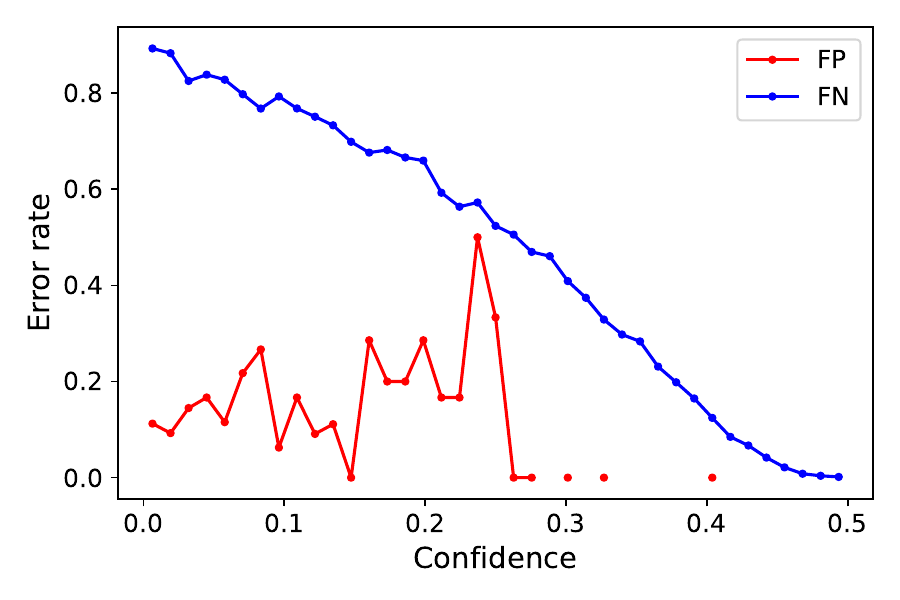}\\
    \caption{Histogram of confidence measures (left column) and error rates for each bin of confidence measures (right column). RandomForest model trained and then validated on Reddit text (top row), arXiv text (middle row), and StackExchange text (bottom row).}
    \label{fig:confidence_validation_RandomForest}
\end{figure}

\begin{figure}[t]
    \centering
    \includegraphics[width=0.49\textwidth]{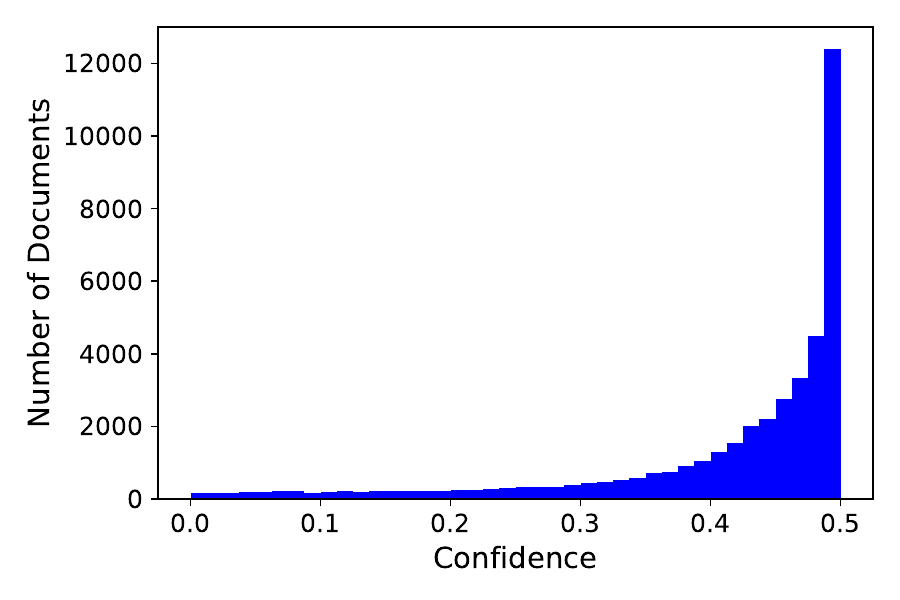}
    \includegraphics[width=0.49\textwidth]{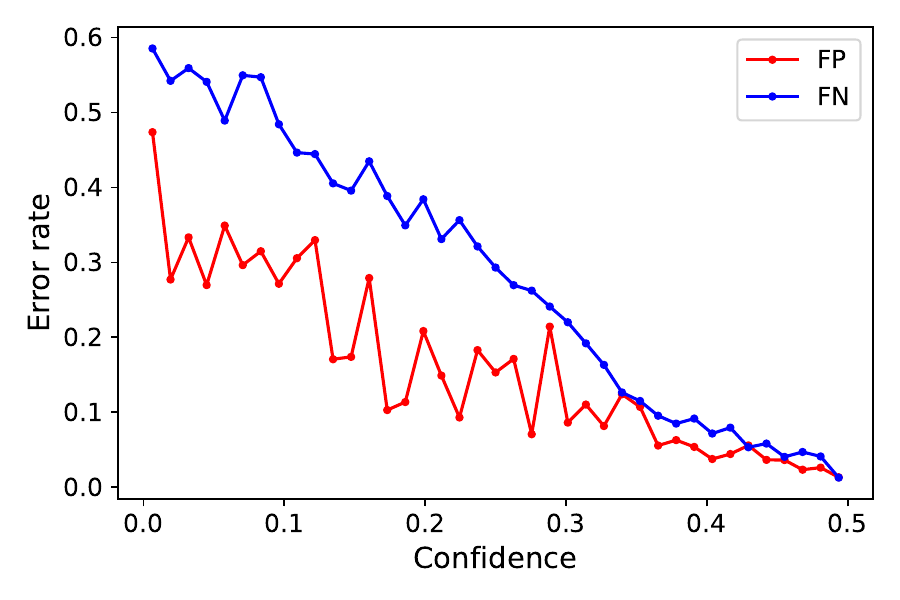}\\
    \includegraphics[width=0.49\textwidth]{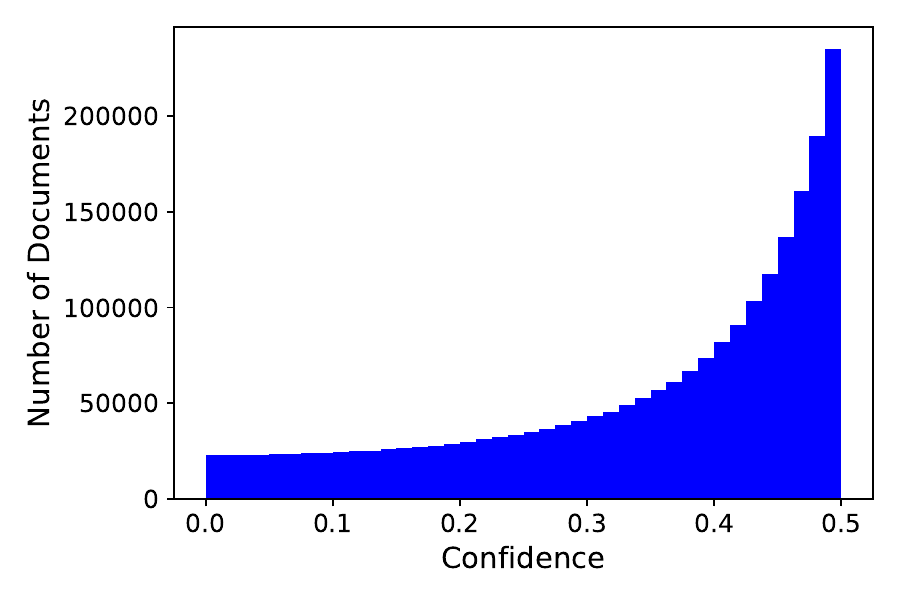}
    \includegraphics[width=0.49\textwidth]{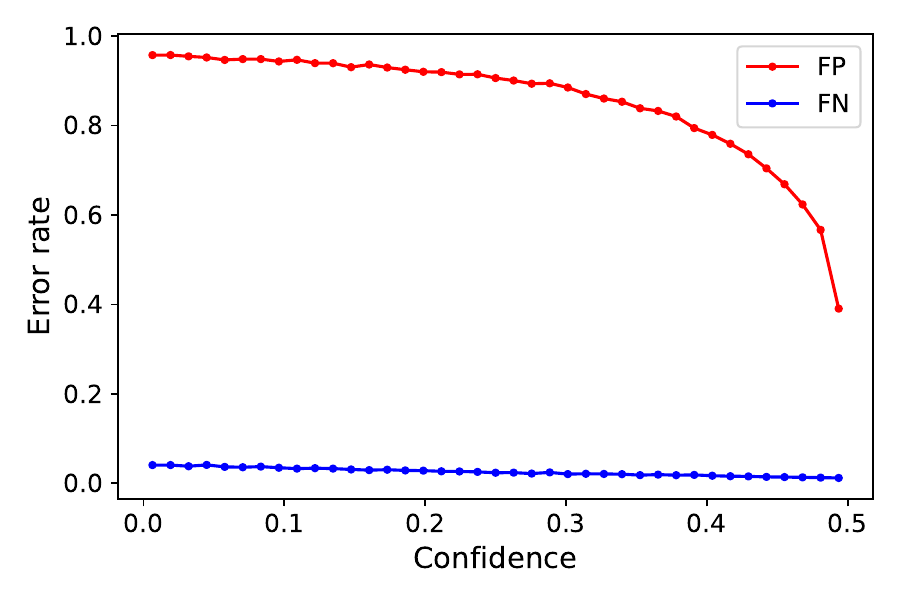}\\
    \caption{Histogram of confidence measures (left column) and error rates for each bin of confidence measures (right column). Logistic model trained on StackExchange text and then validated on arXiv text (top row), Reddit text (bottom row)}
    \label{fig:confidence_cross_validation_stackexchange}
\end{figure}

\begin{figure}[t]
    \centering
    \includegraphics[width=0.49\textwidth]{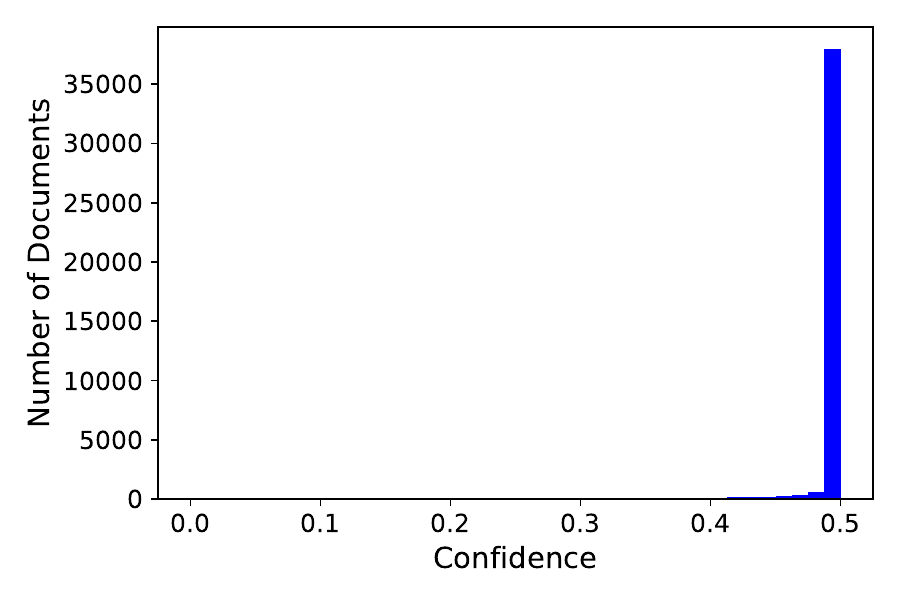}
    \includegraphics[width=0.49\textwidth]{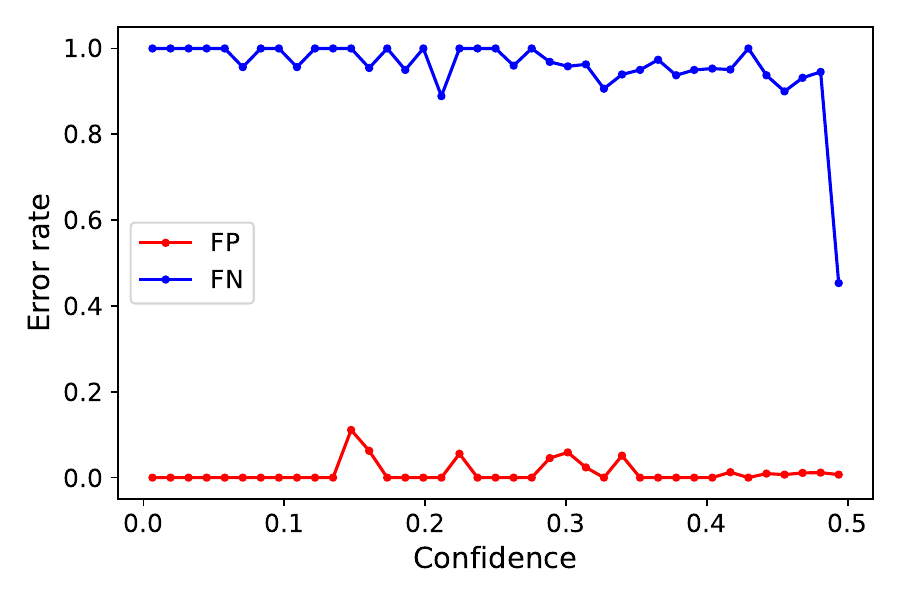}\\
    \includegraphics[width=0.49\textwidth]{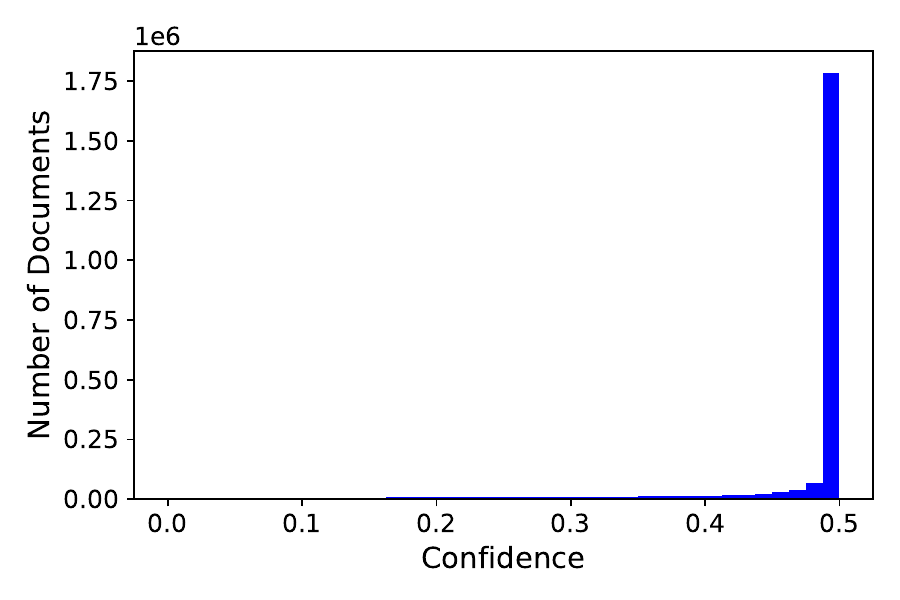}
    \includegraphics[width=0.49\textwidth]{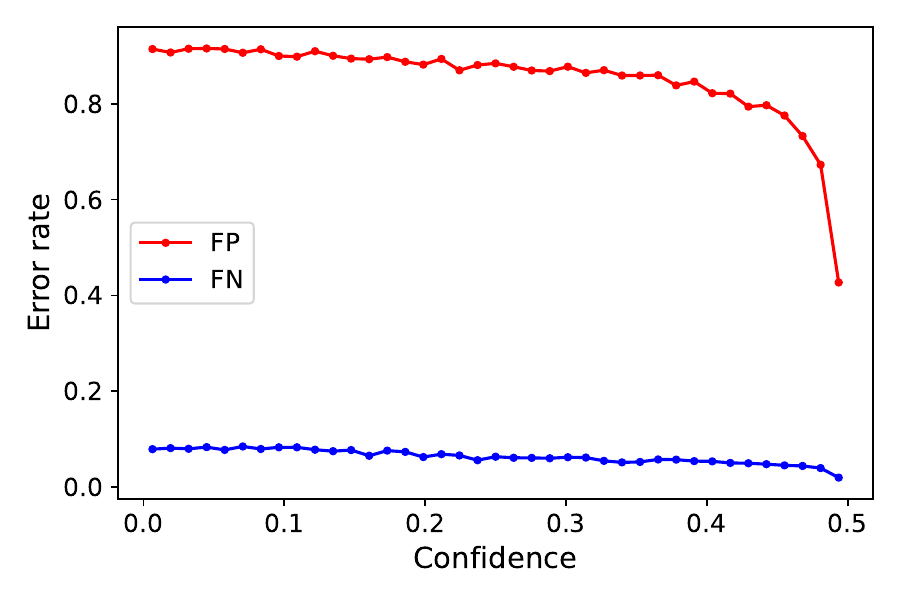}
    \caption{Histogram of confidence measures (left column) and error rates for each bin of confidence measures (right column). DNN $0.95$ model trained on StackExchange text and then validated on arXiv text (top row), Reddit text (bottom row)}
    \label{fig:confidence_cross_validation_stackexchange_DNN0.95}
\end{figure}

Here we define a confidence measure for the continuous output values of the machine learning models. Not all of the machine learning models produce continuous output values, but for the models that do we can define a simple confidence measure that quantifies how far the outputs are from either $0$ or $1$ in the final decision. 

For each sample that is classified (if the model produces continuous output), we get an output vector of the form $[v_0, v_1]$, where $v_0 + v_1 = 1$, $v_0 \geq 0$, and $v_1 \geq 0$. $v_0$ corresponds to the output for the class $0$ (in our system that means not cybersecurity related), and $v_1$ corresponds to the output for the class $1$ (in our system that means the text is cybersecurity related). The confidence measure for a particular document $d$ (with model output $[v_0, v_1]$) is then $conf_d = max([v_0, v_1]) - 0.5$. Thus the $conf_d$ measure is within $[0, 0.5]$, where $0$ indicates no confidence (i.e., the machine learning model effectively predicted a 50-50 coin flip), and $0.5$ is the highest confidence output. A natural question is whether a higher confidence measure from the model corresponds to higher solution quality (i.e., lower false negative and false positive rates). In order to evaluate this, we compute the confidence measures for labelled data in the validation datasets from Section \ref{sec:experiments_validation}, then bin the documents according to their confidence measure and determine the error rates for each of these bins of text samples. This confidence measure could be applied to models in general (with the exception of LinearSVC because the output from LinearSVC is not continuous). However, models such as RandomForest and DecisionTree can produce outputs that are not distributed across the confidence measure range, which can lead to potentially choppy results, where for some region of confidence measures there are no documents predicted in that region. The DNN and Logistic models give consistent continuous output, so these are the models we will demonstrate this analysis on. 

For analysis of these confidence measures, we show two associated figures for each data set and machine learning model analyzed. In the left side graph of each figure, we show the number of documents in each bin versus the confidence measure for the bin. In the right side graph of each figure, we show the error rate for both false positives (FP) and false negatives (FN) of documents in each confidence measure bin. These two figures for each data set and machine learning model show the correlation of our confidence measure with the false positive and false negative error rates. 

Figure's \ref{fig:confidence_validation_Logistic}, \ref{fig:confidence_validation_DNN0.95}, \ref{fig:confidence_validation_DNN0.99}, \ref{fig:confidence_validation_RandomForest}, \ref{fig:confidence_cross_validation_stackexchange}, and \ref{fig:confidence_cross_validation_stackexchange_DNN0.95} show how the error rates (right hand side columns) change for different confidence measure values (left hand side columns). Across all of the error rate plots we observe that both the false negative and false positive rates decrease as confidence gets higher; indicating that higher confidence \emph{does} correspond to lower error rates. Across all of the confidence value histograms, we observe that the model's predict at very high confidence; the one slight exception here is Figure \ref{fig:confidence_validation_Logistic} middle left plot, where the histogram peaks slightly before $0.5$. Note that the instability observed in the arXiv validation dateset plots is due to the small sample size of the arXiv dataset. Conversely, on the StackExchange validation datatsets we observe small overall changes to the error rates, up until very high confidence levels. 

Figures \ref{fig:confidence_validation_DNN0.99} and \ref{fig:confidence_validation_DNN0.95} show the differences between the DNN model trained to $0.99$ and $0.95$ accuracy respectively. The primary differences are that the $0.99$ accuracy trained models have significantly higher confidence values.

\subsection{Cybersecurity Topic Classification (CTC) tool}
\label{sec:CTC}

Now we combine all of these trained models (the validation data for these models was shown in the previous section) into a single NLP tool for cybersecurity topic modeling. In particular, for a given set of documents, we vectorize using the TF-IDF vectorizer used in all of these experiments. Then, we run all 21 models on those documents and report the results. To come to a final decision as the output of the tool, we take the majority vote on the output of the 21 ML models. This is a reasonable approach because it is simply taking the consensus of all of the trained models. This means that no outlier can change the tool's decision making process, making the system robust against outlier predictions.

\subsubsection{CTC Error rates}
On this section we show that the CTC tool on average performs better than any of the individual 21 trained models. 

We use the labelled dataset of \cite{Zhang2015CharacterlevelCN} for this experiment. In particular, we use both the training and testing dataset from \cite{Zhang2015CharacterlevelCN} in this validation test. The dataset has four different labels: 1 (World), 2 (Sports), 3 (Business), 4 (Sci/Tech). Since 4 can include cybersecurity content, we entirely remove 4, and then merge classes 1, 2, and 3 into non cybersecurity. Collectively, we call this data source \textbf{ag-news}. We also use the \textbf{philosophy} \cite{kcalizadeh}, which is definitely not cybersecurity discussion. Both \textbf{ag-news} and \textbf{philosophy} are used as large datasets which provide an idea of the \textit{false positive} rate. 

\begin{figure}[h!]
    \centering
    \includegraphics[width=0.6\textwidth]{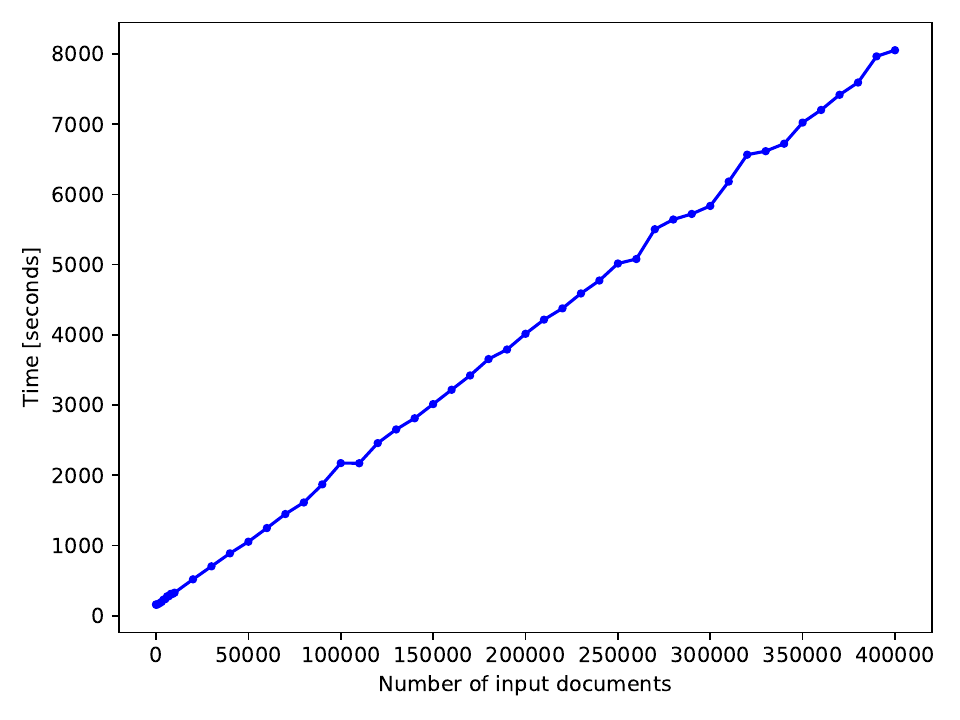}
    \caption{Number of input documents vs Wall clock time for the CTC tool}
    \label{fig:CTC_timing}
\end{figure}

Next, we want to develop a labelled dataset for cybersecurity related text to validate the performance of CTC. To this end, we pull a random subset of 100 documents from seven internet forums/discussion blogs and then hand label the topic's of each of these 700 documents. Note that several of these sources are heavily cybersecurity related, which is valuable as we want to get a reasonable sample of cybersecurity text with which to validate the models. In several cases, the random documents we accessed did not have any English words or enough English words (i.e., some documents had only one or two words), so those documents were discarded. In total, from the seven sources, we have 698 labelled documents.

\begin{figure}[h!]
    \centering
    \includegraphics[width=0.8\textwidth]{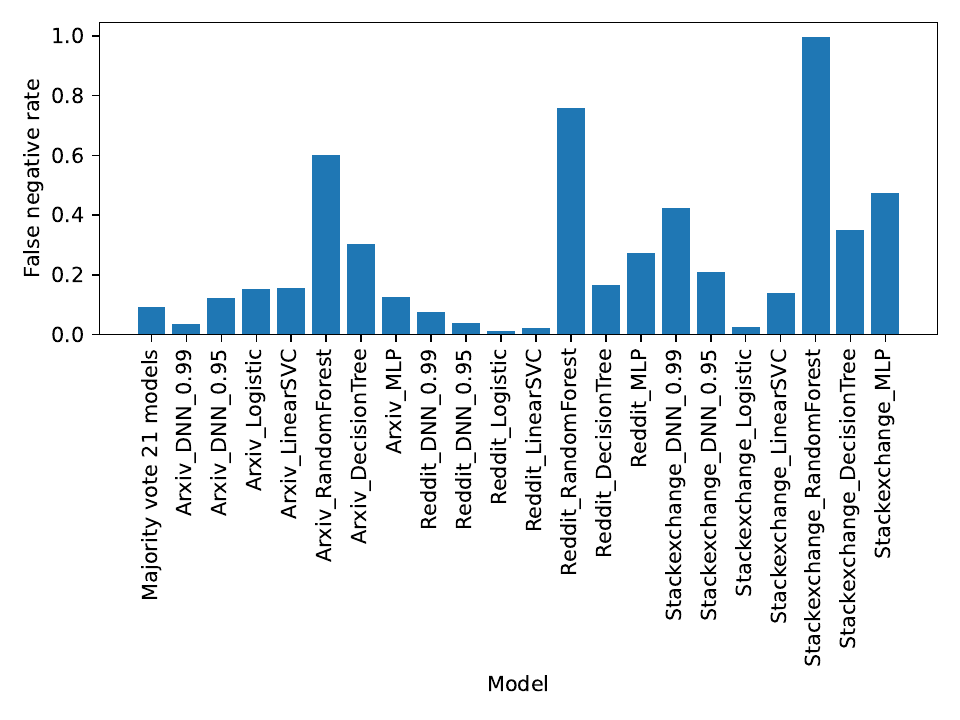}
    \includegraphics[width=0.8\textwidth]{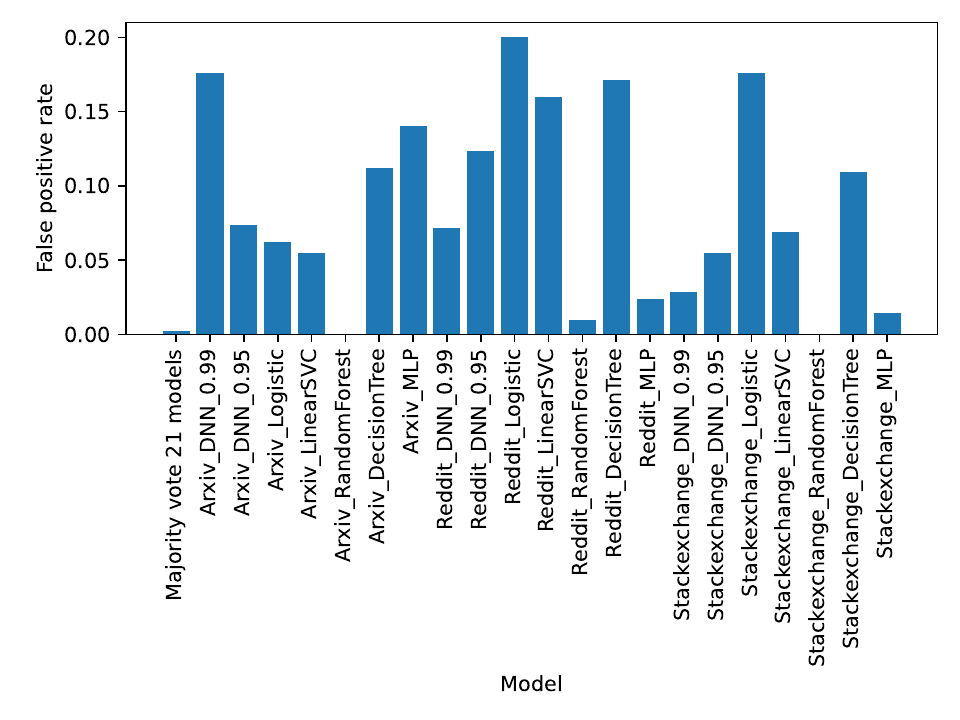}
    \caption{CTC error rate comparison to single ML models. False negative rate comparison across models (top) and False positive rate comparison across models (bottom). }
    \label{fig:CTC_error_rates}
\end{figure}

\begin{table*}[ht]
\centering
\begin{tabular}{|p{4.5cm}||p{1.5cm}|p{1.7cm}|p{5.6cm}|}
    \hline
    Text source and label & Num. of docs. & 21 models incor. lbld. & individual models incorrectly labelled vector\\
    \hline
    \hline
    Google security blog Cybersec & 98 & 3 & [0, 3, 5, 8, 63, 22, 4, 3, 0, 0, 0, 73, 11, 22, 47, 20, 3, 10, 98, 37, 54]\\
    \hline
    Darkreading Cybersec & 98 & 4 & [2, 9, 6, 8, 41, 23, 11, 6, 2, 0, 1, 83, 21, 29, 27, 17, 1, 11, 97, 30, 34]\\
    \hline
    Fireeye Cybersec & 94 & 6 & [2, 11, 5, 6, 56, 27, 3, 3, 3, 2, 2, 71, 10, 18, 51, 20, 3, 19, 94, 30, 52]\\
    \hline
    Ycombinator Cybersec & 2 & 2 & [1, 1, 2, 2, 2, 1, 1, 2, 2, 0, 2, 2, 1, 2, 2, 2, 1, 2, 2, 1, 2]\\
    \hline
    schneier Cybersec & 35 & 5 & [2, 7, 14, 12, 29, 7, 5, 4, 2, 1, 1, 20, 4, 7, 19, 13, 1, 7, 34, 14, 24]\\
    \hline
    toms-forum Cybersec & 1 & 0 & [0, 0, 0, 0, 1, 0, 0, 1, 0, 0, 0, 1, 0, 1, 0, 0, 0, 0, 1, 1, 1]\\
    \hline
    thehackernews Cybersec & 92 & 19 & [8, 20, 32, 30, 60, 47, 28, 13, 7, 2, 3, 69, 23, 35, 32, 16, 2, 10, 92, 34, 32] \\
    \hline
    \hline
    Google security blog Not Cybersec & 2 & 2 & [2, 2, 2, 2, 0, 1, 2, 1, 2, 2, 2, 0, 1, 0, 0, 0, 2, 1, 0, 0, 1]\\
    \hline
    Darkreading Not Cybersec & 1 & 0 & [0, 0, 0, 0, 0, 0, 0, 0, 1, 1, 1, 0, 1, 0, 0, 0, 0, 0, 0, 0, 0]\\
    \hline
    Fireeye Not Cybersec & 5 & 0 & [2, 1, 0, 0, 0, 3, 1, 0, 0, 2, 1, 0, 2, 0, 1, 0, 3, 1, 0, 2, 0]\\
    \hline
    Ycombinator Not Cybersec & 98 & 0 & [22, 3, 2, 0, 0, 2, 14, 7, 14, 22, 15, 0, 19, 0, 1, 2, 22, 6, 0, 5, 0]\\
    \hline
    schneier Not Cybersec & 65 & 15 & [30, 19, 20, 21, 0, 41, 27, 11, 17, 31, 24, 4, 36, 6, 5, 15, 37, 19, 0, 38, 5]\\
    \hline
    toms-forum Not Cybersec & 99 & 0 & [14, 5, 0, 0, 0, 0, 12, 10, 16, 20, 19, 0, 12, 4, 5, 6, 6, 0, 0, 1, 0]\\
    \hline
    thehackernews Not Cybersec & 8 & 0 & [4, 1, 2, 0, 0, 0, 3, 1, 2, 6, 5, 0, 1, 0, 0, 0, 4, 2, 0, 0, 0]\\
    \hline
    \hline
    ag-news Not Cybersec & 95698 & 521 & [13957, 8143, 1119, 1670, 0, 3832, 12603, 285, 862, 6435, 4565, 104, 4920, 345, 939, 1638, 14158, 9228, 112, 5929, 1319] \\
    \hline
    philosophy Not Cybersec & 360808 & 434 & [37654, 23525, 284, 454, 0, 1074, 35416, 34027, 52652, 13267, 23344, 59, 4831, 3913, 79093, 68829, 23065, 27708, 12, 2631, 4992] \\
    \hline
\hline
\end{tabular}
\caption{Number of incorrectly labelled documents across multiple text sources. The first column shows the labelled text source. Second column shows the total number of documents from that text source. The third column shows the total number of documents that the CTC tool (i.e., the majority vote of the 21 models) incorrectly labelled. The fourth column shows the number of incorrectly labelled documents by each of the 21 individual models. }
\label{table:CTC_validation}
\end{table*}

Table \ref{table:CTC_validation} shows the number of incorrectly labelled documents when using the CTC tool (the majority vote of the 21 individual models) and when using each of the individual 21 models applied to the new validation data described above. On average, the CTC tool outperforms the individual models across different text sources. Figure \ref{fig:CTC_error_rates} shows a more concise version of Table \ref{table:CTC_validation}, where we aggregate the results across the \textit{cybersecurity} and \textit{non cybersecurity} labelled validation text. Figure \ref{fig:CTC_error_rates} shows that while there are some individual models that have very low \textit{false positive} \emph{or} low \textit{false negative} rates, the majority vote of the 21 models has the lowest overall \textit{false positive} \emph{and} low \textit{false negative}. For example, we observe that \textit{Arxiv-RandomForest} has a very low false positive rate, but then has a very high false negative rate. Thus, the majority vote mechanism used in the CTC tool is more robust compared to the individual models. 

\subsubsection{CTC Confidence Metric}
\begin{figure}[t]
    \centering
    \includegraphics[width=0.49\textwidth]{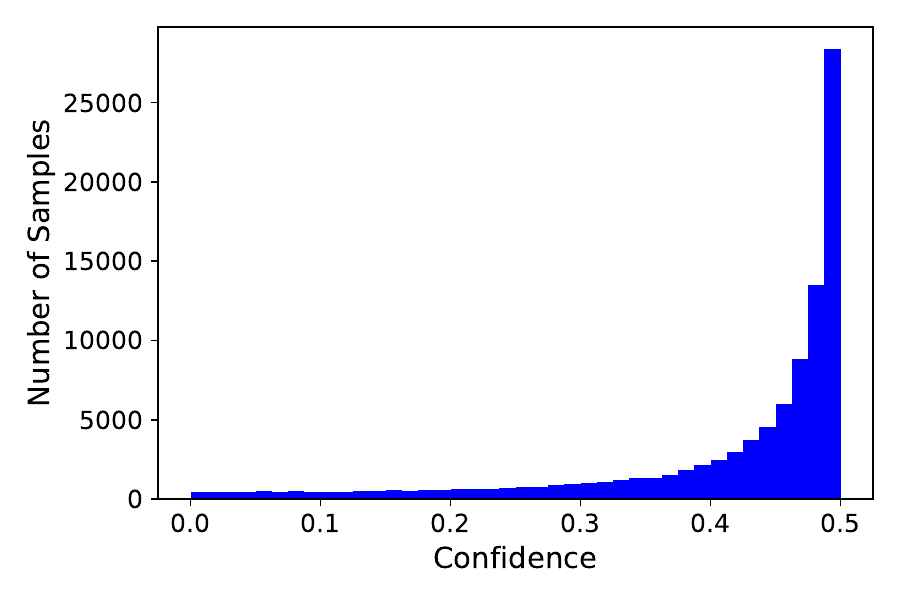}
    \includegraphics[width=0.49\textwidth]{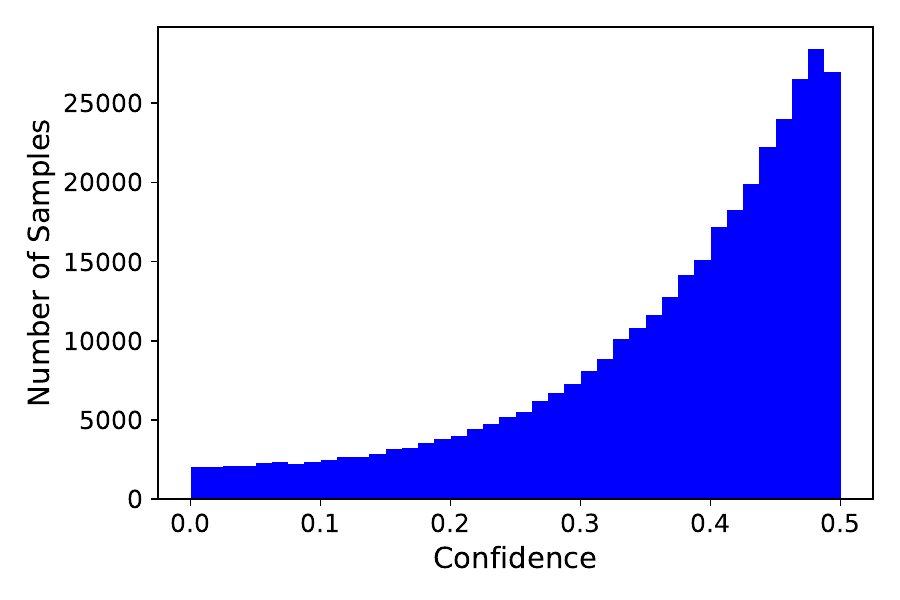}\\
    \includegraphics[width=0.49\textwidth]{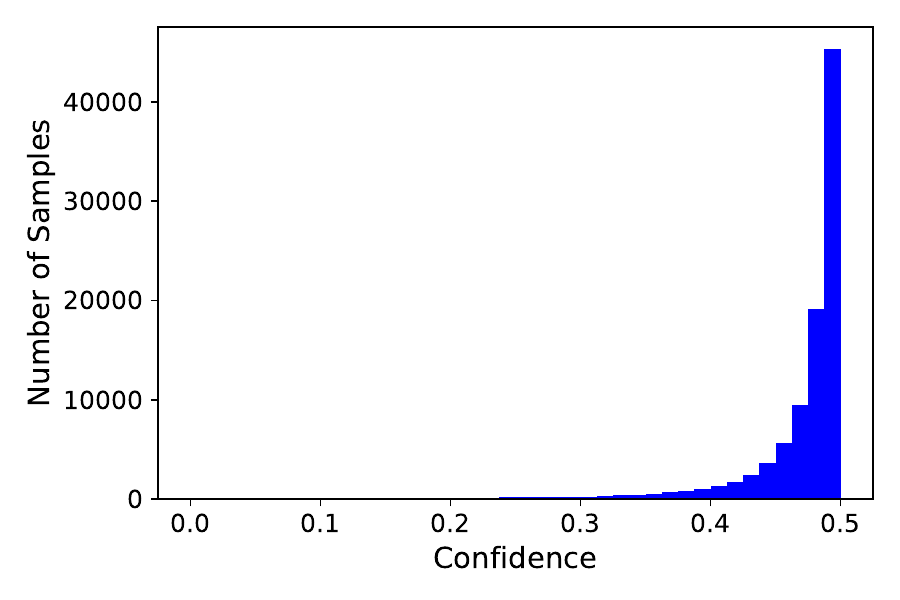}
    \includegraphics[width=0.49\textwidth]{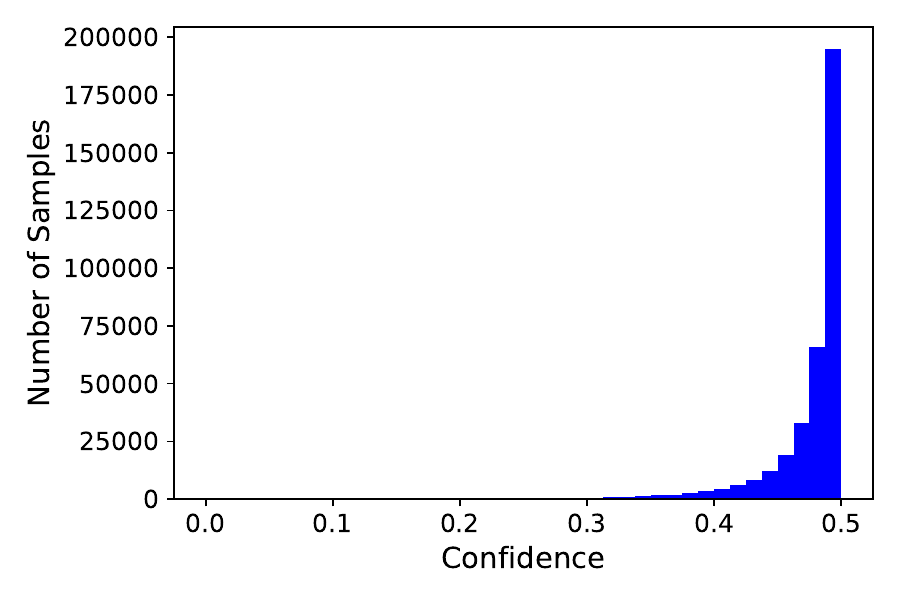}\\
    \includegraphics[width=0.49\textwidth]{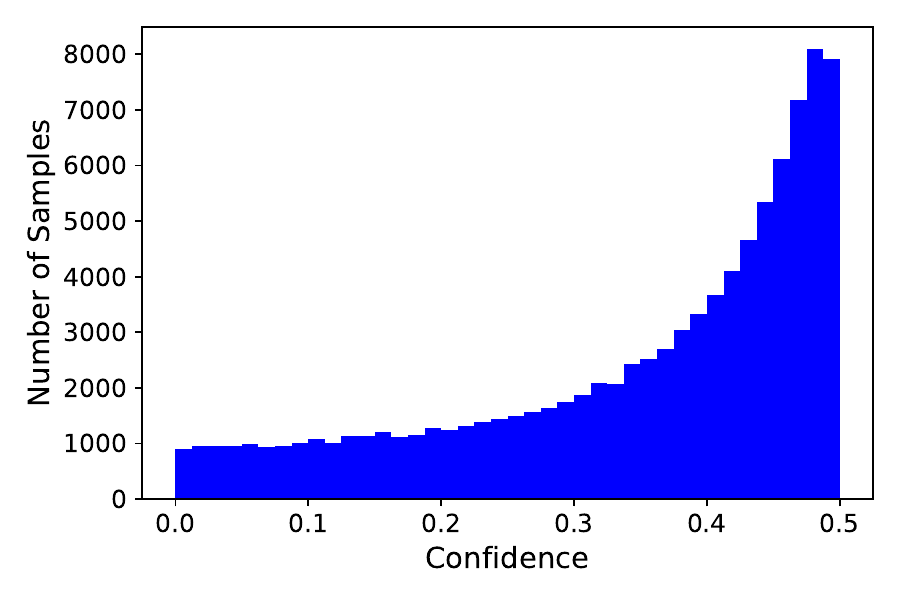}
    \includegraphics[width=0.49\textwidth]{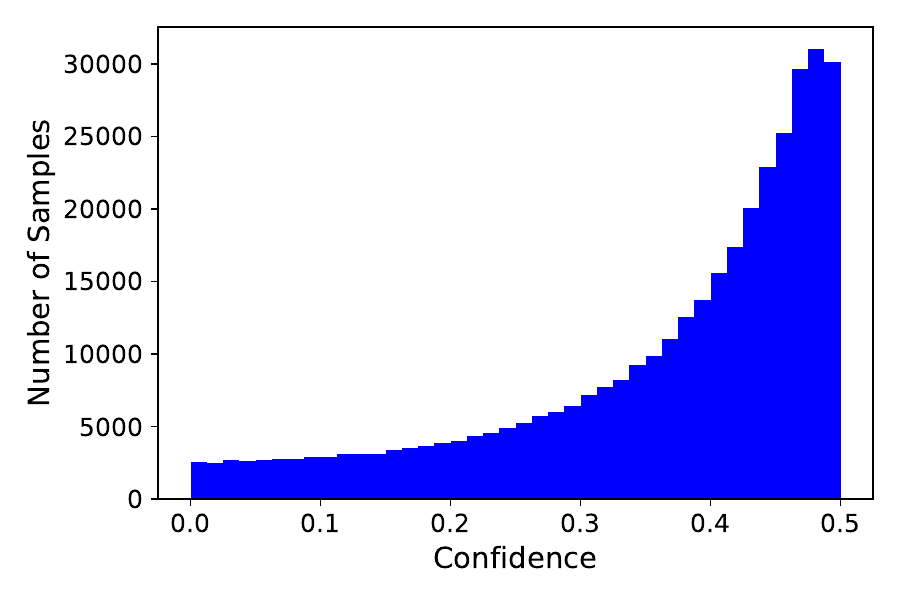}
    \caption{Histogram of confidence measures from the Logistic model trained on Reddit text (top row), arXiv text (middle row), and StackExchange text (bottom row) for the \textbf{ag-news} (left column) and \textbf{philosophy} (right column) datasets.}
    \label{fig:CTC_confidence_validation_Logistic}
\end{figure}

The models that make up the CTC tool can still provide confidence measures for the overall decision made by the tool. As an example, Figure \ref{fig:CTC_confidence_validation_Logistic} shows the distribution of confidence values for the three Logistic models in the CTC tool when applied to the \textbf{philosophy} (right column) dataset and the \textbf{ag-news} dataset (left column). We see that the distribution of confidence values changes depending on the data the model was trained on, as well as the data the model is predicting on. In particular, the StackExchange trained models (bottom row) have less confident outputs compared to the arXiv trained models (middle row). The Reddit trained models (top row) have more confident outputs when classifying the \textbf{ag-news} dataset compared to the \textbf{philosphy} dataset. The behavior of the Reddit text trained model is different than the other two models in this respect; this could be because the Reddit text is predominantly very short text length (see Figure \ref{fig:Reddit_token_length_histogram}), whereas the other two source had higher diversity of text length. As with the previous confidence measure plots, we observe that model's predict at very high confidence (meaning very close to the maximum confidence measure of $0.5$) on average. 

\subsubsection{CTC Timing}
Lastly, we characterize how the CTC tool scales in terms of documents analyzed over time. We pull random subsets of some internet discussion posts of varying length and content and then measure the total wall clock time needed to classify that set of documents. We repeat this process for increasing numbers of input documents. 

Figure \ref{fig:CTC_timing} shows this scaling of documents analyzed over time. We observe that the wall clock time needed to classify $N$ input documents has a consistently linear scaling. This scaling is largely due to using multiprocessing for the Tensorflow DNN models. The limiting factor of how many documents CTC can ingest is the available RAM on the host computer. For the result shown in Figure \ref{fig:CTC_timing}, the device had 500 GB of RAM, but we do not characterize exactly where the RAM limit is.

\section{Conclusion and Future Work}
\label{sec:conclusion}

This article proposed a methodology for gathering labelled English text for the purpose of topic modelling cybersecurity discussions. We then trained multiple machine learning models using this labelled data, and showed that combining these models into a consensus majority voting tool results in both low error rates and good scalability in terms of text samples classified per hour. 

There remain many possible future research avenues:

\begin{enumerate}[noitemsep]
    \item Considering unsupervised clustering methods, which can determine how similar each source of training data is, and therefore also how similar a new input document is to each of these sources can yield lower error rates and lower computation times by using only a single machine learning classifier that has the highest accuracy for that type of document (based on the validation experiments done previously). 
    \item Using clustering methods to first cluster the corpus of labelled training data into a large number of clusters and then separately training machine learning models on each of those clusters could yield higher accuracy when using all of the models together. 
    \item Using other vectorization algorithms that make use of a larger dictionary of words, as well as the sequence in which words are used, may improve performance over using a bag of words model.
    \item Gathering more labelled data from other online sources, for example from Twitter, Quroa, or Medium, would increase the size of the training set. These sites also offer methods of mining labelled natural language posts, for example \cite{DBLP:conf/aaai/LippmanWMCCS017} and \cite{10.1145/3132847.3132866} use data from Twitter. 
    \item Analyzing the explainability of the machine learning algorithms could generate new insight. For example, the decision tree algorithm might provide a more compact and explainable logical decision path for this particular topic modeling task. 
    \item Using other machine learning algorithms that may be better suited to topic classification tasks, for example Recurrent Neural Networks. 
    \item Exploring the use of subset's of the 21 machine learning models (with the majority voting mechanism) could improve results and reduce computational time. In particular, some subset may perform better than all 21 models and individual models. 
    \item Including non English text may provide additional valuable data once translated. 
\end{enumerate}

\section*{Acknowledgements}
Sandia National Laboratories (SNL) is a multi-program laboratory managed and operated by Sandia Corporation, a wholly owned subsidiary of Lockheed Martin Corporation, for the U.S. Department of Energy's National Nuclear Security Administration under contract DE-AC04-94AL85000. The New Mexico Cybersecurity Center of Excellence (NMCCoE) is a statewide Research and Public Service Project supported center for economic development, education, and research. The authors would like to thank both SNL and NMCCoE for funding and computing system access/support.

\setlength\bibitemsep{0pt}
\printbibliography

\end{document}